\documentclass[usegraphicx,usenatbib]{mn2e}
\usepackage{amsmath}
\usepackage{amssymb}
\usepackage{times}
\usepackage{subfigure}
\usepackage{epsfig}
\usepackage{multirow}

\renewcommand{\descriptionlabel}[1]%
  {\hspace{\labelsep}\textbf{#1}}

\title[Systematic Trends In SDSS Photometric Data]
      {Systematic Trends In Sloan Digital Sky Survey Photometric Data}

\author[D. M. Bramich and W. Freudling]
  {D. M. Bramich$^{1}$\thanks{E-mail: dbramich@eso.org, dan.bramich@hotmail.co.uk}  \medskip
   and W. Freudling$^{1}$ \medskip
  \\$^{1}$European Southern Observatory, Karl-Schwarzschild-Stra{\ss}e 2, 85748 Garching bei M\"{u}nchen, Germany
  }

\begin{document}

\date{Accepted 2011 August ???. Received 2011 August ???; Submitted 2011 August ???}

\pagerange{\pageref{firstpage}--\pageref{lastpage}} \pubyear{2010}

\maketitle

\label{firstpage}

\begin{abstract}
We investigate the Sloan Digital Sky Survey (SDSS) photometry from Data Release 8 (DR8)
in the search for systematic trends that still exist after the calibration effort of Padmanabhan et al.
We consider both the aperture and point-spread function (PSF) magnitudes in DR8.
Using the objects with repeat observations, we find that a large proportion of the aperture magnitudes
suffer a $\sim$0.2-2\% systematic trend as a function of PSF full-width half-maximum (FWHM), the amplitude of which increases
for fainter objects. Analysis of the PSF magnitudes reveals more complicated systematic trends
of similar amplitude as a function of PSF FWHM and object brightness.
We suspect that sky over-subtraction is the cause of the largest amplitude trends as a function of PSF FWHM.
We also detect systematic trends as a function of subpixel coordinates for the PSF magnitudes
with peak-to-peak amplitudes of $\sim$1.6~mmag and $\sim$4-7~mmag for the over- and under-sampled
images, respectively. We note that the systematic trends are similar in amplitude to the reported $\sim$1\% 
and $\sim$2\% precision of the SDSS photometry in the $griz$ and $u$ wavebands, respectively, and therefore their correction has the potential to
substantially improve the SDSS photometric precision. We provide an {\tt IDL} program specifically for this purpose.
Finally, we note that the SDSS aperture and PSF magnitude scales are related by a non-linear transformation
that departs from linearity by $\sim$1-4\%, which, without correction, invalidates the application
of a photometric calibration model derived from the aperture magnitudes to the PSF magnitudes, as has
been done for SDSS DR8.
\end{abstract}


\section{Introduction}
\label{sec:intro}

The SDSS (\citealt{yor2000}) provides photometry (including widely-used aperture and PSF magnitudes)
for objects down to $r\sim$22.5~mag and covering 14555~deg$^{2}$ (\citealt{aih2011} - from now on AIH11). The photometric
calibration is uniform over the survey area at the $\sim$1\% and $\sim$2\% levels in the $griz$ and $u$ wavebands, respectively
(\citealt{pad2008} - from now on PAD08). This vast
data set has provided insight into many areas of astronomical research, including the large-scale structure of the Universe, properties of galaxies, Galactic
structure, stellar populations, just to name a few.

For many of the scientific applications of SDSS photometric data, an accurate absolute photometric calibration (i.e. knowledge of fluxes in physical units such as 
J~m$^{-2}$~s$^{-1}$) is not as important as a precise relative calibration over the survey. Furthermore, it is substantially more
challenging to achieve an accurate absolute calibration due to the inherent difficulties in minimising the systematic uncertainties in the mapping between the spectral
energy distribution (SED) of an appropriate (and usually very bright) fundamental spectrophotometric standard and the natural photometric system of the survey.
For these reasons, much recent effort has been invested in determining a precise relative calibration of SDSS imaging data,
and other similar surveys have followed suit (e.g. \citealt{reg2009}). Future ground-based all-sky surveys that require 1\% or better relative photometric calibrations
(e.g. PanSTARRS - \citealt{kai2002}, LSST - \citealt{ive2008}, etc.) are investing considerable resources to perform
exquisitely precise measurements of the telescope-instrument-detector system throughput as a function of wavelength for each imager pixel
(\citealt{stu2006}; \citealt{stu2010}) and to use spectroscopic observations of calibrator stars coupled with detailed atmospheric modelling to enable the precise correction of
atmospheric extinction effects to the level of a few millimagnitudes (\citealt{stu2007}; \citealt{bur2010}).

The purpose of a relative photometric calibration is to model the systematic differences in the measured magnitudes of survey objects as they are observed
over time and over the spatial extent of the survey. The sources of these differences are multitude although they are usually dominated by atmospheric
extinction variations, differences in detector responses, and errors introduced during the photometric analysis of the images. Ideally, correction
of the full survey data using the fitted calibration model should result in object photometry that reaches some target homogeneity such as the desirable 1\% level.
The key to being able to fit a relative photometric calibration across survey data is the existence of {\it repeat observations} of a sufficient number of
objects to link all the observational data via the adopted calibration model and to provide strong enough constraints on the model parameters.
The early development of the photometric calibration methodology may be found in the works of \citet{har1981}, \citet{ree1982}, \citet{man1983},
\citet{hon1992}, and \citet{man1995}.

PAD08 present the impressive development of a photometric calibration model for SDSS data on a scale many times larger than has been 
used before. The model is fit to the repeat observations from the overlaps between the scan columns and at the ends of the scans along great circles
(see Section~\ref{sec:camera}), and from the repeat scanning of certain scan regions. For SDSS Data Release 6 (\citealt{ade2008}), their model
has $\sim$2000 calibration parameters of interest and $\sim$10$^{7}$ nuisance parameters (the star mean magnitudes) to be solved for.
PAD08 marginalise over the unknown star mean magnitudes and iteratively solve the normal equations
of the smaller least squares problem for the calibration parameters. 
The resulting relative calibration is good to $\sim$1\% in the $griz$ wavebands ($\sim$2\% in the $u$ waveband).

In PAD08, the authors looked at the calibration residuals as a function of star brightness (or mean magnitude), as a function of detector column to assess the
accuracy of the flat-field calibrations, as a function of time, and as a function of celestial coordinates, the latter being further decomposed into spatial error modes.
They found a clear trend as a function of time, which manifests as ``coherent errors at the few millimagnitude level'', and which
most likely correlates with time-varying image quality parameters such as seeing, sky brightness, etc. An example of 
such a time-dependent trend may be discerned as a $\sim$10~mmag drift in the magnitude residuals in the upper panel of Fig.~9 in PAD08.
This implies that the PAD08 photometric model may need to be updated to include appropriate terms to fit these trends.
We also highlight the fact that the photometric calibration parameters in PAD08 are derived from fitting the SDSS aperture magnitudes and the calibration is then applied to 
calibrate various object magnitude measurements in the survey catalogues such as PSF magnitudes.

The purpose of this paper is to extend the investigation into the systematic trends in the SDSS photometric data. We stress
that an exhaustive study of this topic is virtually impossible and not within the scope of our work.
We start in Section~\ref{sec:modelling} by describing our modelling method. In Section~\ref{sec:photo_data}, we describe the procedure we use
to select appropriate photometric data from the SDSS data base. We then investigate
the systematic trends in the SDSS aperture and PSF magnitude measurements, and we report on the trends we find as a function of
PSF FWHM, object brightness, and subpixel coordinates (Section~\ref{sec:systrend}). In Section~\ref{sec:compare}, we derive the relation
between the aperture and PSF magnitude scales, which we find suffers from non-linearities. We summarise and discuss
our results in Section~\ref{sec:disc}.

\section{Modelling The Photometric Data}
\label{sec:modelling}

\subsection{The Photometric Model}
\label{sec:phot_model}

Our approach to identifying the systematic trends in the SDSS photometric data consists of
investigating how the photometry depends on each object or image property which we suspect could
have an impact on the photometry. Our aim is to identify any significant trends that were not modelled
during the PAD08 calibration of the aperture magnitudes. However, since we do not know {\it a priori} the
form of any potential trend, we cannot assume a smooth functional form at this stage.

Instead, we introduce a binning for an object/image property $X$ that covers the full range of data values that we are modelling.
For the $k$th bin in $X$, we introduce an unknown magnitude offset $Z_{k}$ to be determined,
the purpose of which is to model the mean difference of the photometric measurements within the corresponding bin in $X$ from the rest of the photometric
measurements\footnote{Note that a {\it positive} magnitude offset $Z_{k}$ indicates that the photometric measurements   
within the corresponding bin in $X$ are {\it fainter} on average than the rest of the photometric measurements, and vice
versa.}. Our photometric model may then be written as:
\begin{equation}
\overline{m}_{i} = \sum_{p = 1}^{N_{\mbox{\tiny obj}}} \delta_{jp} \, M_{p} + \sum_{p = 1}^{N_{\mbox{\tiny z}}} \delta_{kp} \, Z_{p} = M_{j} + Z_{k}
\label{eqn:phot_model}
\end{equation}
where $\overline{m}_{i}$ is the model magnitude for the $i$th magnitude measurement $m_{i}$, $M_{j}$ is the unknown true instrumental magnitude of
the $j$th object, $N_{\mbox{\scriptsize obj}}$ is the number of objects in the photometric data sample,
$N_{\mbox{\scriptsize z}}$ is the number of magnitude offsets $Z_{k}$ that we are attempting to fit, and $\delta_{ij}$ is the Kronecker delta-function:
\begin{equation}
\delta_{ij} =   
\begin{cases}
1 & \mbox{if $i = j$} \\  
0 & \mbox{if $i \ne j$} \\
\end{cases}
\label{eqn:kronecker_delta}
\end{equation}
Note that the $i$th photometric measurement in our photometric data
sample belongs to the $j(i)$th object and the $k(i)$th bin in $X$, where the adopted notation for $j$ and $k$ reflects the
fact that both of the indices $j$ and $k$ are functions of the index $i$. However, in the rest of this paper, we devolve to using the notation $j$ and $k$ for
$j(i)$ and $k(i)$, respectively, in order to avoid confusion in our subscript notation.

The photometric model defined in Equation~\ref{eqn:phot_model} is degenerate since increasing each of the instrumental magnitudes $M_{j}$
by an arbitrary amount $c$ at the same time as decreasing each of the magnitude offsets $Z_{k}$ also by $c$ will have no effect
on the model values $\overline{m}_{i}$. This degeneracy may be removed by fixing the instrumental magnitude
of a single object to an arbitrary but convenient value. This is equivalent to fixing the absolute photometric calibration of the observations.

\subsection{Constructing The Normal Equations}
\label{sec:normal_equations}

Equation~\ref{eqn:phot_model} is linear and therefore we may use general linear least-squares (\citealt{pre2007})
to find the solution for the instrumental magnitudes $M_{j}$ and the magnitude offsets $Z_{k}$. The chi-squared associated
with Equation~\ref{eqn:phot_model} is:
\begin{equation}
\chi^{2} = \sum_{i = 1}^{N_{\mbox{\tiny data}}} \left( \frac{m_{i} - \sum_{p = 1}^{N_{\mbox{\tiny obj}}} \delta_{jp} \, M_{p}
                                                - \sum_{p = 1}^{N_{\mbox{\tiny z}}} \delta_{kp} \, Z_{p}}{\sigma_{i}} \right)^{2}
\label{eqn:chi2}
\end{equation}
where $N_{\mbox{\scriptsize data}}$ is the number of photometric measurements, and $\sigma_{i}$ is the uncertainty on the $i$th photometric measurement.

The normal equations of the least-squares problem are found by differentiating the $\chi^{2}$ in Equation~\ref{eqn:chi2}
with respect to each parameter in the model and they may be written in matrix form as:
\begin{equation}
\left( \begin{matrix} \textbf{A} & \textbf{B} \\ \textbf{B$^{\textbf{T}}$} & \textbf{D} \end{matrix} \right)
\left( \begin{matrix} \textbf{x$_{1}$} \\ \textbf{x$_{2}$} \end{matrix} \right)
= \left( \begin{matrix} \textbf{v$_{1}$} \\ \textbf{v$_{2}$} \end{matrix} \right)
\label{eqn:normal_eqns}
\end{equation}
where \textbf{A}, \textbf{B}, and \textbf{D} are matrices of sizes $N_{\mbox{\scriptsize obj}} \times N_{\mbox{\scriptsize obj}}$,
$N_{\mbox{\scriptsize obj}} \times N_{\mbox{\scriptsize z}}$, and $N_{\mbox{\scriptsize z}} \times N_{\mbox{\scriptsize z}}$ elements,
respectively, and where \textbf{v}$_{1}$ and \textbf{v}$_{2}$ are vectors of lengths $N_{\mbox{\scriptsize obj}}$ and $N_{\mbox{\scriptsize z}}$
elements, respectively. The vector \textbf{x}$_{1}$ contains the $N_{\mbox{\scriptsize obj}}$ unknown instrumental magnitudes $M_{j}$, and
the vector \textbf{x}$_{2}$ contains the $N_{\mbox{\scriptsize z}}$ unknown magnitude offsets $Z_{k}$.
The individual elements of \textbf{A}, \textbf{B}, \textbf{D}, \textbf{v}$_{1}$, and \textbf{v}$_{2}$ are given by:
\begin{eqnarray}
A_{pq}  & = & \sum_{i = 1}^{N_{\mbox{\tiny data}}} \delta_{jp} \, \delta_{jq} \, / \, \sigma_{i}^2 \label{eqn:Ael} \\
B_{pq}  & = & \sum_{i = 1}^{N_{\mbox{\tiny data}}} \delta_{jp} \, \delta_{kq} \, / \, \sigma_{i}^2 \label{eqn:Bel} \\
D_{pq}  & = & \sum_{i = 1}^{N_{\mbox{\tiny data}}} \delta_{kp} \, \delta_{kq} \, / \, \sigma_{i}^2 \label{eqn:Del} \\
v_{1,p} & = & \sum_{i = 1}^{N_{\mbox{\tiny data}}} \delta_{jp} \, m_{i} \, / \, \sigma_{i}^2   \label{eqn:v1el} \\
v_{2,p} & = & \sum_{i = 1}^{N_{\mbox{\tiny data}}} \delta_{kp} \, m_{i} \, / \, \sigma_{i}^2   \label{eqn:v2el}
\end{eqnarray}

Matrix \textbf{A} (Equation~\ref{eqn:Ael}) is diagonal since the product $\delta_{jp} \, \delta_{jq}$ is only
non-zero when $j = p = q$, which makes sense because a photometric measurement $m_{i}$ may only correspond to a single object.
Similarly, matrix \textbf{D} (Equation~\ref{eqn:Del}) is diagonal, which is due to the fact that a photometric measurement $m_{i}$
may only correspond to a single bin in $X$. Matrix \textbf{B} (Equation~\ref{eqn:Bel}) happens to be sparse when there are
generally less photometric measurements per object than bins in $X$.
Finally, we observe that the elements of the vectors \textbf{v}$_{1}$ and \textbf{v}$_{2}$ (Equations~\ref{eqn:v1el}~\&~\ref{eqn:v2el}) are
simply the inverse-variance weighted sums of the photometric measurements $m_{i}$ for each object and for each bin in $X$, respectively.

At this stage, it is worth noting that the construction of the least squares matrix and the vector on the right-hand side is a highly
parallel computational problem due to the natural partitioning of the photometric measurements on a per-object and per-bin-in-$X$ basis,
which is a very convenient fact because this is where a large fraction of the computational operations occur in finding the solution for \textbf{x}$_{1}$
and \textbf{x}$_{2}$. Furthermore, the facts that \textbf{A} and \textbf{D} are diagonal, and \textbf{B} is sparse, may be used to implement
computer code that makes efficient use of available memory resources, especially when it may be unfeasible to store the full versions of
the matrices \textbf{A}, \textbf{B}, and \textbf{D} in computer memory.

\subsection{Solving The Normal Equations}
\label{sec:solving_normal_equations}

The fact that \textbf{A} is diagonal in the normal equations for photometric models including the true instrumental magnitudes
(as in Equation~\ref{eqn:phot_model}) has been spotted previously by \citet{reg2009}, and it is clearly the case in \citet{hon1992} even
though no comment was made. In Appendix~A of \citet{reg2009}, they
use this property to develop a tractable way of solving the normal equations for the parameter vectors \textbf{x}$_{1}$
and \textbf{x}$_{2}$, and for calculating the marginalised covariance matrix \textbf{C} for the parameters in \textbf{x}$_{2}$.
We briefly repeat this method here using the notation in this paper.

Eliminating the parameter vector \textbf{x}$_{1}$ from the normal equations in Equation~\ref{eqn:normal_eqns} yields:
\begin{equation}
\left( \textbf{D} - \textbf{B$^{\textbf{T}}$} \textbf{A}^{-1} \textbf{B} \right) \textbf{x}_{2}
= \textbf{v}_{2} - \textbf{B$^{\textbf{T}}$} \textbf{A}^{-1} \textbf{v}_{1}
\label{eqn:x2sol}
\end{equation}
where the elements of \textbf{A}$^{-1}$ are easily computed as:
\begin{equation}
\textbf{A}^{-1}_{pq} =
\begin{cases}
1/A_{pq} & \mbox{if $p = q$} \\
0        & \mbox{if $p \ne q$} \\
\end{cases}
\label{eqn:Ainv}
\end{equation}
Cholesky factorisation of the symmetric and positive-definite\footnote{We do not provide the proof of this statement in this paper.}
matrix $\left( \textbf{D} - \textbf{B$^{\textbf{T}}$} \textbf{A}^{-1} \textbf{B} \right)$, followed by forward and back substitution
is the most efficient and numerically stable method (\citealt{gol1996}) for obtaining the solution for the magnitude offsets $Z_{k}$. The solution for
the instrumental magnitudes $M_{j}$, if required, can be
obtained by substituting the solution for \textbf{x}$_{2}$ into Equation~\ref{eqn:normal_eqns}:
\begin{equation}
\textbf{x}_{1} = \textbf{A}^{-1} \textbf{v}_{1} - \textbf{A}^{-1} \textbf{B} \textbf{x}_{2}
\label{eqn:x1sol}
\end{equation}
Finally, the marginalised covariance matrix \textbf{C} for the magnitude offsets $Z_{k}$ may be obtained by marginalising over the
true instrumental magnitudes:
\begin{equation}
\textbf{C} = \left( \textbf{D} - \textbf{B$^{\textbf{T}}$} \textbf{A}^{-1} \textbf{B} \right)^{-1}
\label{eqn:covar}
\end{equation}

\subsection{Implementation And Processing Time}
\label{sec:exec_time}

We have implemented an {\tt IDL} program called {\tt fit\_photometric\_calibration.pro} as part of the DanIDL\footnote{http://www.danidl.co.uk} library
of routines. This program allows the user to fit a fully configurable and highly flexible photometric calibration model (including static and rotating
illumination corrections - \citealt{moe2010}) to a set of magnitude
measurements, and it employs the methodology developed in Sections~\ref{sec:normal_equations}~\&~\ref{sec:solving_normal_equations}.
The specific photometric model defined in Equation~\ref{eqn:phot_model} uses only one of a slew of available terms that may be
configured in the photometric model in {\tt fit\_photometric\_calibration.pro}.

The program that we have developed works solely by storing all the necessary data and arrays in computer memory, which limits the size of the
fitting problem that can be tackled to the amount of memory that is available. If we improved our program to capitalise on the highly parallel
nature of the computational problem by reading and writing data from the computer hard disk as necessary, then the limit on the size of the fitting
problem that can be solved can be greatly increased to match with the amount of disk space available since computer hard disks currently obtain much larger
storage sizes than computer memory. By doing this, we would also remove any constraint imposed by the amount of computer memory available, because the program
would only use a limited and well-defined amount of memory at any one time. However, for the analysis of the data presented in this paper, our
{\tt IDL} program is not limited by the amount of computer memory that is available.

We have briefly tested the performance of our program and its scalability to larger data sets. Using a single Intel Xeon 2.0Ghz CPU on a 64-bit machine with 132Gb of RAM,
we find that for $N_{z} = 75$ and $N_{\mbox{\scriptsize data}} = 3 N_{\mbox{\scriptsize obj}}$, our program uses $\sim$1.7, 4.4, and 8.9~Gb of RAM for
$N_{\mbox{\scriptsize obj}} = 0.6\times10^{6}$, $1.5\times10^{6}$, and $3\times10^{6}$, respectively, while taking $\sim$1.6, 3.8, and 8.5~minutes to run, respectively.
The processing limit for our program on the same machine seems to lie at approximately $N_{\mbox{\scriptsize obj}} = 10^{7}$,
$N_{\mbox{\scriptsize data}} = 10^{8}$, and $N_{z} = 100$, which requires $\sim$48~Gb of RAM and takes $\sim$52~minutes to run.

\section{SDSS Photometric Data}
\label{sec:photo_data}

\subsection{The SDSS Imaging Camera}
\label{sec:camera}

The SDSS imaging camera, mounted on a dedicated 2.5m telescope at the Apache Point Observatory (New Mexico), consists
of an array of 30 SITe/Tektronix charge-coupled devices (CCDs), each of size 2048$\times$2048 pix and with
a pixel scale of 0.396\arcsec~pix$^{-1}$, arranged in six columns of
five chips each with a space of approximately one chip width between columns (\citealt{gun1998}; \citealt{gun2006}).
Each row of six chips is positioned behind a different filter so that SDSS imaging data are produced in five wavebands, namely
$u$, $g$, $r$, $i$, and $z$ (\citealt{fuk1996}; \citealt{smi2002}). The camera operates in time-delay-and-integrate (TDI)
readout mode scanning along great circles in the sky at the sidereal rate (\citealt{zar1996}). The camera scan
direction is parallel to the six columns of CCDs, and the temporal observation order of the wavebands
is $riuzg$. The chip arrangement is such that two scans may be used to cover a filled stripe 2\fdg54 wide,
with $\sim$1\arcmin ($\sim$8\%) overlap between chip columns in the two scans. The telescope and camera optics, along with the
CCD layout, are designed such that, given perfect tracking along great circles, the optical distortion over the field-of-view (FOV) results in 
star tracks across each CCD that deviate from being parallel to the CCD columns by at most $\sim$0.06\arcsec (or $\sim$0.15~pix) over
the length of a detector in the worst case.

\subsection{Data Selection From SDSS DR8}
\label{sec:data}

We obtained our photometric data sample from SDSS DR8 (AIH11), which includes all imaging data up to the
retirement of the SDSS camera, reprocessed with an updated version of the SDSS photometric pipeline, and calibrated using the PAD08
modelling scheme. We employed the CasJobs interface\footnote{http://skyservice.pha.jhu.edu/casjobs/} to query the DR8 data base.
We extracted photometric observations of all stars\footnote{In reality, we are selecting PSF-like
point-source objects which may include quasars, asteroids, etc.}
(TYPE$=$6) that have no ``child objects'' (NCHILD$=$0; \citealt{sto2002})
and calibrated (CALIBSTATUS\_FILTER AND 1) PSF magnitudes brighter than 19~mag (PSFMAG\_FILTER$<$19) with
positive uncertainties smaller than 1~mag (0$<$PSFMAGERR\_FILTER$<$1). The magnitude threshold
was chosen to limit our data sample to those objects with photometric uncertainties of better than $\sim$2-3\%.
The photometric measurements are supplied with corresponding quality flags which aid in the selection of ``good quality'' measurements. We
further filtered our data by applying the set of quality constraints described in Table~2 of \citet{bra2008}.
Finally, the resolution of the individual objects to which each photometric observation belongs is available in DR8 and the
object identifications are stored in the THINGID entry in the DR8 data base. Where this identification process fails, a THINGID value
of $-$1 is stored, and we therefore dropped such measurements from our data sample.

\begin{figure*}
\centering
\begin{tabular}{cc}
\subfigure[]{\epsfig{file=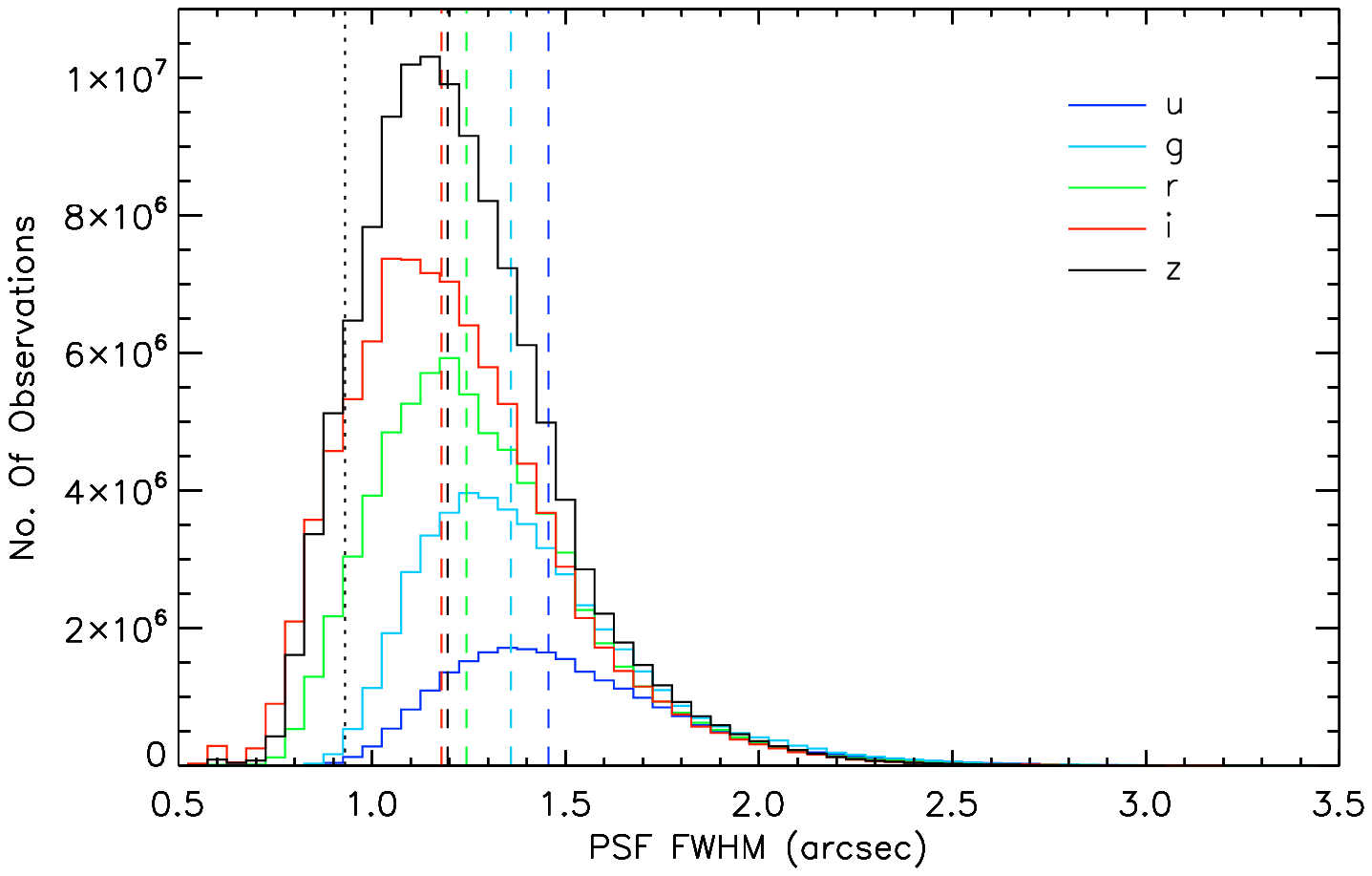,angle=0.0,width=0.5\linewidth} \label{fig:hist_fwhm_dr8}} &
\subfigure[]{\epsfig{file=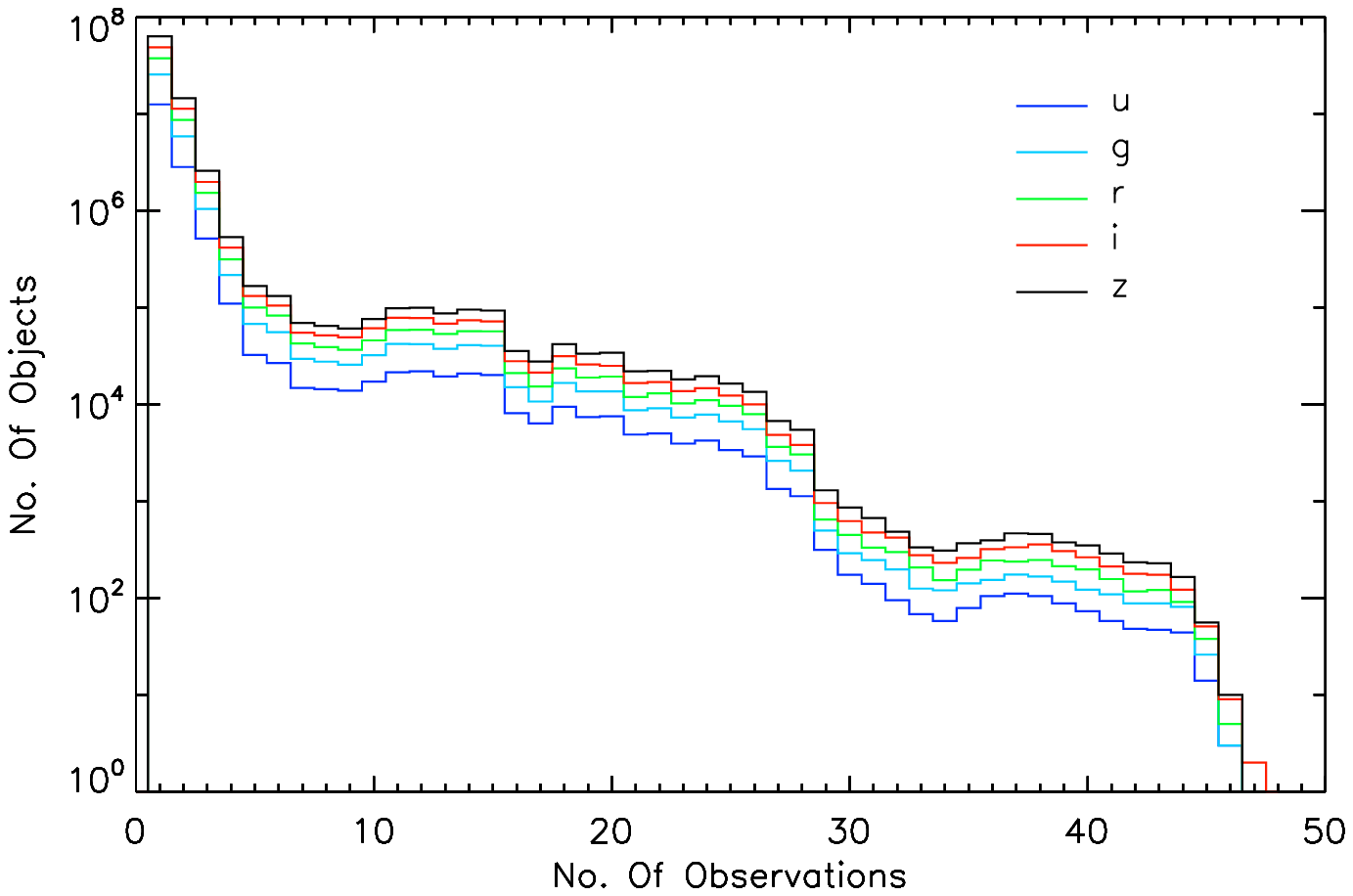,angle=0.0,width=0.5\linewidth} \label{fig:hist_nep_dr8}} \\
\end{tabular}
\caption{(a) Histograms of PSF FWHM (arcsec) over the object observations for each filter, using data for the six camera columns combined. The vertical dashed lines
             represent the median PSF FWHMs for each filter, and the vertical dotted black line represents the PSF FWHM corresponding to critical
             sampling (0.93\arcsec) for the SDSS imaging detectors.
         (b) Histograms of the number of observations of each object over the set of objects for each filter, using data for the six camera columns combined.
         Both: The data used in these plots are the observations we selected from SDSS DR8 in Section~\ref{sec:data}.
         \label{fig:histograms_dr8}}
\end{figure*}

\begin{table*}
\caption{The properties of the selected observations from SDSS DR8 (see Section~\ref{sec:data}) for each
         of the 30 SDSS imaging detectors. The data sets for each detector are organised by filter and
         camera column. The 5th, 50th (median), and 95th percentile values of the cumulative distribution of PSF FWHM values
         (arcsec) are listed in columns 7, 8, and 9, respectively.
         }
\centering
\begin{tabular}{ccccccccc}
\hline
Filter & Camera & \multicolumn{2}{c}{PSF Magnitudes}   & \multicolumn{2}{c}{Aperture Magnitudes} & \multicolumn{3}{c}{PSF FWHM (arcsec)}     \\
       & Column & No. Of Objects & No. Of Observations & No. Of Objects & No. Of Observations    & 5th Percentile & Median & 95th Percentile \\
\hline
$u$    & 1      & 2552954        & 3680087             & 2540345        & 3661050                & 1.21           & 1.56   & 2.19            \\
$u$    & 2      & 2717648        & 3970893             & 2698799        & 3941161                & 1.17           & 1.50   & 2.15            \\
$u$    & 3      & 2694623        & 3944556             & 2678855        & 3920169                & 1.05           & 1.40   & 2.07            \\
$u$    & 4      & 2743806        & 4020514             & 2726917        & 3994001                & 1.04           & 1.39   & 2.05            \\
$u$    & 5      & 2719071        & 3964543             & 2702696        & 3939132                & 1.05           & 1.40   & 2.05            \\
\smallskip
$u$    & 6      & 2786132        & 4092273             & 2768708        & 4064715                & 1.13           & 1.47   & 2.10            \\
$g$    & 1      & 5511948        & 7946945             & 5432614        & 7835474                & 1.07           & 1.39   & 2.00            \\
$g$    & 2      & 5515143        & 7978389             & 5429156        & 7856395                & 1.06           & 1.36   & 1.97            \\
$g$    & 3      & 5457676        & 7889987             & 5383166        & 7783855                & 1.03           & 1.34   & 1.94            \\
$g$    & 4      & 5434436        & 7859519             & 5360404        & 7753636                & 1.02           & 1.33   & 1.93            \\
$g$    & 5      & 5632190        & 8169352             & 5552972        & 8056549                & 1.03           & 1.35   & 1.95            \\
\smallskip
$g$    & 6      & 5608587        & 8139232             & 5535674        & 8036950                & 1.06           & 1.38   & 1.97            \\
$r$    & 1      & 8142785        & 11698216            & 7492919        & 10718162               & 0.94           & 1.27   & 1.83            \\
$r$    & 2      & 8061099        & 11595390            & 7377343        & 10558970               & 0.90           & 1.23   & 1.79            \\
$r$    & 3      & 8067250        & 11635364            & 7384481        & 10558611               & 0.87           & 1.18   & 1.76            \\
$r$    & 4      & 8103852        & 11695080            & 7434713        & 10668356               & 0.88           & 1.19   & 1.75            \\
$r$    & 5      & 7761595        & 11210722            & 7092689        & 10191515               & 0.93           & 1.24   & 1.80            \\
\smallskip
$r$    & 6      & 8409320        & 12184076            & 7770240        & 11218480               & 1.01           & 1.34   & 1.87            \\
$i$    & 1      & 10424397       & 14944150            & 9689614        & 13844569               & 0.88           & 1.21   & 1.75            \\
$i$    & 2      & 10737931       & 15545164            & 9959987        & 14360505               & 0.82           & 1.13   & 1.68            \\
$i$    & 3      & 10198773       & 14609342            & 9484867        & 13532858               & 0.80           & 1.10   & 1.65            \\
$i$    & 4      & 10236481       & 14778349            & 9519267        & 13683502               & 0.81           & 1.12   & 1.66            \\
$i$    & 5      & 11060536       & 16060059            & 10274455       & 14865032               & 0.86           & 1.16   & 1.70            \\
\smallskip
$i$    & 6      & 10712588       & 15516036            & 9991929        & 14435648               & 0.99           & 1.33   & 1.84            \\
$z$    & 1      & 13596773       & 19575551            & 11964994       & 17056492               & 0.96           & 1.27   & 1.77            \\
$z$    & 2      & 13679274       & 19774085            & 11953310       & 17082888               & 0.88           & 1.17   & 1.68            \\
$z$    & 3      & 13745026       & 19896381            & 11997206       & 17169875               & 0.83           & 1.11   & 1.62            \\
$z$    & 4      & 13922390       & 20149555            & 12046553       & 17227888               & 0.85           & 1.13   & 1.63            \\
$z$    & 5      & 13593815       & 19575105            & 11946407       & 17040156               & 0.88           & 1.18   & 1.68            \\
$z$    & 6      & 13359060       & 19139276            & 11765172       & 16715773               & 0.97           & 1.32   & 1.80            \\
\hline
\end{tabular}
\label{tab:dr8_data}
\end{table*}

The aperture photometry measurements in the SDSS DR8 data base are stored as fluxes in units of ``nanomaggies'', defined such that an object of brightness 1~nanomaggie
has a magnitude of 22.5 in any waveband. In our data sample, we included the aperture photometry measurements for the commonly used ``aperture-7'' aperture of
radius 7.43\arcsec, and we converted these flux measurements and their uncertainties into standard SDSS ``asinh magnitudes''
using the equations in \citet{lup1999} and the standard SDSS values for the ``softening'' parameter $b$ of 0.14, 0.09,
0.12, 0.18, and 0.74~nanomaggies in $ugriz$, respectively. We note that our data are selected on the
basis of PSF magnitudes, and sometimes the SDSS pipeline failed to measure a corresponding aperture flux.
We flagged these cases for later exclusion when we wish to perform calculations using the aperture magnitudes.

We also performed a neighbourhood search within our data for each waveband independently, recording for each object the number of objects within
a radius of 15\arcsec, and storing this information with each photometric observation in our data set.

Our final data set consists of 30 data tables, one for each detector, listing for each object observation the object identification, the
right ascension and declination, the aperture and PSF magnitude measurements and their uncertainties, the PSF FWHM, the detector $x$ and $y$ coordinates,
and the number of objects from our data set within 15\arcsec. In Table~\ref{tab:dr8_data}, we report the number of objects and the number of
associated photometric observations in our data set for each detector. These quantities differ between the PSF and aperture magnitude
measurements due to the failure of the SDSS pipeline to measure an aperture flux in some cases.

In Fig.~\ref{fig:hist_fwhm_dr8}, we plot a histogram of the PSF FWHM (arcsec) over the object observations for each filter.
The vertical dashed lines represent the median PSF FWHMs of $\sim$1.46\arcsec, 1.36\arcsec, 1.24\arcsec, 1.18\arcsec, and 1.20\arcsec$\,$
for the $ugriz$ wavebands, respectively. We note that the actual PSF FWHM distributions for each detector  
among detectors with the same filter differ at the $\sim$5-20\% level due to PSF variations across the camera FOV,
with the detectors in the outer camera columns having a larger PSF FWHM than those in the inner camera columns.
This may be seen in columns 7, 8, and 9 of Table~\ref{tab:dr8_data} where we list the 5th, 50th (median), and 95th percentile values
of the cumulative distribution of PSF FWHM values for each detector, respectively.

In Fig.~\ref{fig:hist_nep_dr8}, we plot a histogram of the number of observations of each object over the set of objects for each filter.
Clearly most objects have only a single observation in SDSS DR8. Those objects that do have repeat observations generally have only two or three repeat observations,
highlighting the fact that in SDSS DR8, it is the overlaps between scan columns and at the ends of the
scans along great circles that dominate the generation of repeat observations.

\subsection{Repeat Observations}
\label{sec:repeat}

In the absence of external information about the relative brightnesses of objects, it is only
possible to analyse the systematic trends present in a set of photometric measurements by using those objects for which repeat
observations have been performed. An object with a single photometric observation simply provides one data point at the expense of generating one unknown (the
object brightness), bringing no new information to the problem. Hence, for our analysis of the systematic trends present in the SDSS
aperture and PSF magnitudes, we must select the subset of repeat observations from our full DR8 data set.

We work with the data for each detector independently. For the aperture magnitude measurements, we select the
observations for which the SDSS pipeline successfully measured an aperture flux, and then further restrict our selection to those
observations for which an object has at least two observations. We again produce 30 data tables, one for each detector, and in
Table~\ref{tab:dr8_data_repeat_ap}, we report 
the number of objects and the number of associated photometric observations in this data set for each detector.

For the PSF magnitudes, again we select the observations for which an object has at least two observations,
and we produce another 30 data tables, one for each detector. In Table~\ref{tab:dr8_data_repeat_psf}, we report the number of objects and the number of associated 
photometric observations in this data set for each detector.

We note that the PSF FWHM distribution of the repeat observations is very similar to the PSF FWHM distribution of all selected DR8 observations
displayed in Fig.~\ref{fig:hist_fwhm_dr8}, which means that by using the subset of repeat observations to investigate the systematic
trends in the SDSS photometric data, we are not introducing a bias into our results.

\begin{table*}
\caption{The properties of the selected repeat observations with aperture magnitude measurements from SDSS DR8 (see Section~\ref{sec:repeat}) for each
         of the 30 SDSS imaging detectors. The data sets for each detector are organised by filter and camera column. In columns 5-7, we list
         the 4$\sigma$ upper limits $L_{\mbox{\scriptsize under}}$, $L_{\mbox{\scriptsize normal}}$, and $L_{\mbox{\scriptsize well}}$
         that we determine on the existence of
         systematic photometric trends as a function of subpixel coordinates for the regimes of under-sampled, normally-sampled, and well-sampled PSFs,
         respectively. Upper limits are not reported for the $r$ waveband because we actually detect a systematic trend in this case
         (see Section~\ref{sec:apermag3}). In column 8, we report the amount by which the $\chi^{2}$ for the repeat observations
         decreases after calibrating the aperture magnitudes using the fitted magnitude offsets from Section~\ref{sec:apermag2}. In column 9, we
         list the $\Delta \chi^{2}$ values in units of $\sigma$.
         }
\centering
\begin{tabular}{ccccccccc}
\hline
Filter & Camera & No. Of  & No. Of       & $L_{\mbox{\scriptsize under}}$ & $L_{\mbox{\scriptsize normal}}$ & $L_{\mbox{\scriptsize well}}$ & $\Delta \chi^{2}$ & $\Delta \chi^{2} / \sqrt{2 ( N_{\mbox{\scriptsize data}} - N_{\mbox{\scriptsize obj}} )}$ \\
       & Column & Objects & Observations & (mmag)                         & (mmag)                          & (mmag)                        &                   & \\
\hline
$u$    & 1      &  564388 & 1685093      & $-$                            & 0.31                            & 0.21                          &  32089            &   21.4 \\
$u$    & 2      &  624071 & 1866433      & $-$                            & 0.22                            & 0.18                          &  41546            &   26.4 \\
$u$    & 3      &  619495 & 1860809      & 2.47                           & 0.16                            & 0.17                          &  62286            &   39.5 \\
$u$    & 4      &  634159 & 1901243      & 2.16                           & 0.16                            & 0.18                          &  49695            &   31.2 \\
$u$    & 5      &  618910 & 1855436      & 3.17                           & 0.17                            & 0.19                          &  44119            &   28.1 \\
\smallskip
$u$    & 6      &  645631 & 1941638      & $-$                            & 0.18                            & 0.19                          &  32709            &   20.3 \\
$g$    & 1      & 1248996 & 3651856      & 2.24                           & 0.10                            & 0.12                          & 174564            &   79.6 \\
$g$    & 2      & 1254214 & 3681453      & 2.20                           & 0.08                            & 0.10                          & 240578            &  109.2 \\
$g$    & 3      & 1238595 & 3639284      & 1.07                           & 0.08                            & 0.11                          & 192931            &   88.0 \\
$g$    & 4      & 1235798 & 3629030      & 1.01                           & 0.08                            & 0.11                          & 191978            &   87.7 \\
$g$    & 5      & 1288367 & 3791944      & 1.04                           & 0.08                            & 0.11                          & 216183            &   96.6 \\
\smallskip
$g$    & 6      & 1284853 & 3786129      & 2.01                           & 0.09                            & 0.11                          & 129869            &   58.1 \\
$r$    & 1      & 1677458 & 4902701      & N/A                            & N/A                             & N/A                           & 304379            &  119.8 \\
$r$    & 2      & 1647326 & 4828953      & N/A                            & N/A                             & N/A                           & 224208            &   88.9 \\
$r$    & 3      & 1659526 & 4863656      & N/A                            & N/A                             & N/A                           & 187477            &   74.1 \\
$r$    & 4      & 1672184 & 4905827      & N/A                            & N/A                             & N/A                           & 210433            &   82.7 \\
$r$    & 5      & 1596979 & 4695805      & N/A                            & N/A                             & N/A                           & 234588            &   94.2 \\
\smallskip
$r$    & 6      & 1770292 & 5218532      & N/A                            & N/A                             & N/A                           & 198267            &   75.5 \\
$i$    & 1      & 2145815 & 6300770      & 0.27                           & 0.08                            & 0.15                          & 216213            &   75.0 \\
$i$    & 2      & 2264189 & 6664707      & 0.16                           & 0.08                            & 0.15                          & 247950            &   83.6 \\
$i$    & 3      & 2089237 & 6137228      & 0.15                           & 0.08                            & 0.17                          & 210707            &   74.1 \\
$i$    & 4      & 2147740 & 6311975      & 0.16                           & 0.08                            & 0.17                          & 238117            &   82.5 \\
$i$    & 5      & 2350388 & 6940965      & 0.21                           & 0.07                            & 0.15                          & 213343            &   70.4 \\
\smallskip
$i$    & 6      & 2272797 & 6716516      & 0.93                           & 0.10                            & 0.15                          & 167352            &   56.1 \\
$z$    & 1      & 2609860 & 7701358      & 0.71                           & 0.11                            & 0.20                          & 195523            &   61.3 \\
$z$    & 2      & 2621322 & 7750900      & 0.33                           & 0.10                            & 0.21                          & 218830            &   68.3 \\
$z$    & 3      & 2641188 & 7813857      & 0.23                           & 0.10                            & 0.23                          & 199848            &   62.1 \\
$z$    & 4      & 2647421 & 7828756      & 0.29                           & 0.11                            & 0.25                          & 182612            &   56.7 \\
$z$    & 5      & 2594648 & 7688397      & 0.40                           & 0.11                            & 0.23                          & 144198            &   45.2 \\
$z$    & 6      & 2520684 & 7471285      & 0.84                           & 0.14                            & 0.21                          & 172169            &   54.7 \\
\hline
\end{tabular}
\label{tab:dr8_data_repeat_ap}
\end{table*}

\section{Systematic Photometric Trends}
\label{sec:systrend}

\subsection{Aperture Magnitudes}
\label{sec:apermag1}

We may now investigate how the DR8 aperture photometry correlates with various object/image properties to see if
there are any significant systematic trends that were not modelled during the PAD08 calibration of the aperture magnitudes.
Even though the aperture magnitudes themselves are rarely used for high precision photometry, this investigation is crucial
because the fit of the PAD08 calibration model to the aperture magnitudes is used to calibrate various object magnitudes,
including the PSF magnitudes, which are widely used for high precision photometry.
We note that we cannot investigate the existence of systematic trends as a function of object brightness since we are using the
photometry of objects of supposedly constant brightness, and such observations do not contain any information on the magnitude scale itself.
In other words, if there are any non-linearities in the aperture magnitude scale, we cannot detect them using the repeat observations.

We use the data set corresponding to Table~\ref{tab:dr8_data_repeat_ap} and we analyse the photometric data for each
detector independently. For each quantity $X$ that we wish to investigate, we fit the photometric model described in
Section~\ref{sec:phot_model} to the aperture magnitude measurements using an
iterative procedure where we reject photometric observations that lie more than 3$\sigma$ away from the fitted model. We then drop the object
observations for which an object now has only a single observation.
This data rejection step is necessary because of the presence of variable sources and outlier photometric measurements in our data set, and typically
$\sim$3-5\% of the data are rejected at this stage. We then repeat the fit of the photometric model to the cleaned aperture magnitude
measurements.

Applying this iterative fitting procedure to the quantity $X$ representing the detector row with bins of size 10~pix, we find that there are no significant trends in the
fitted magnitude offsets ($Z_{k} < 0.2$~mmag). Also, for the quantity $X$ representing the detector column with bins of size 10~pix,
we find that there are no significant trends in the magnitude offsets ($Z_{k} < 0.5$~mmag).
However, we do find systematic trends in the magnitude offsets when we perform the fit for the quantity $X$ representing
the PSF FWHM.

\subsubsection{Trends As A Function Of PSF FWHM}
\label{sec:apermag2}

For each detector, we partition our data into bins of width 0.1\arcsec$\,$ in PSF FWHM, and we further subdivide the data in each
bin into those observations which have magnitude measurements brighter than 16~mag, in the range 16-18~mag,
and fainter than 18~mag. We construct our photometric model as in Section~\ref{sec:phot_model} adopting an
unknown magnitude offset for each bin, and we follow the iterated fitting procedure described in Section~\ref{sec:apermag1}.

\begin{figure*}
\centering
\epsfig{file=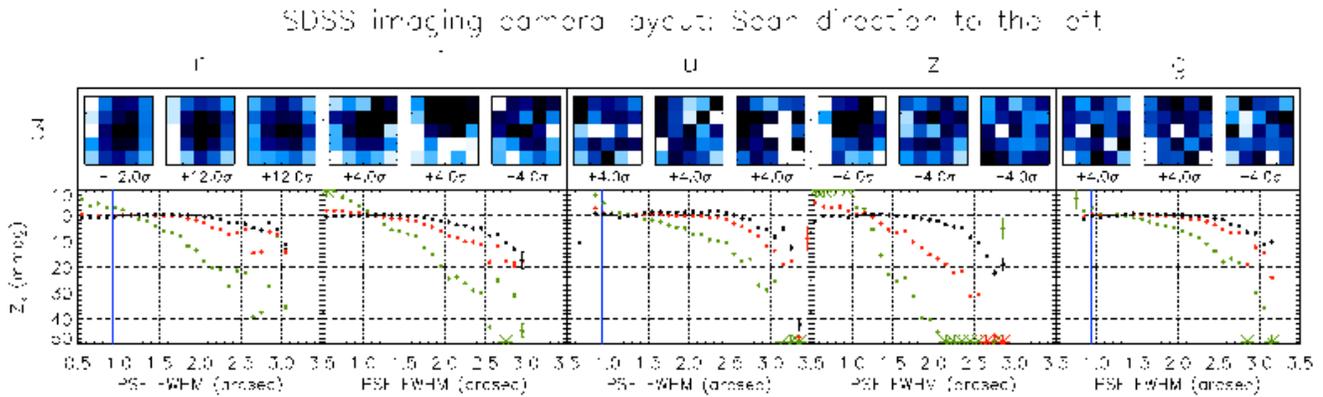,angle=0.0,width=0.99\linewidth}
\caption{Systematic trends in the fitted magnitude offsets $Z_{k}$ for the aperture magnitude measurements from camera column~3
         as a function of PSF FWHM, object brightness, and subpixel coordinates
         (see text in Sections~\ref{sec:apermag2}~\&~\ref{sec:apermag3} for details).
         The data used to generate these plots are the selected repeat observations with aperture magnitude measurements
         from SDSS DR8 (see Section~\ref{sec:repeat}).
         The black, red, and green points correspond to aperture magnitude measurements
         brighter than 16~mag, in the range 16-18~mag, and fainter than 18~mag, respectively.
         Asterisks represent magnitude offsets that fall outside of the plot range. 
         The three images above each plot represent the IPSMs
         for the regimes of under-sampled, normally-sampled, and well-sampled PSFs (from left to right).
         The colour-scale is linear and runs between $\mp$4$\sigma$ and $\mp$12$\sigma$ for the $ugiz$
         and $r$ wavebands, respectively. The full version of this figure including all of the camera columns is available
         from the authors on request.
         \label{fig:maps_ap}}
\end{figure*}

In Fig.~\ref{fig:maps_ap}, for camera column 3, we plot the fitted magnitude offsets for the magnitude measurements brighter than
16~mag as a function of PSF FWHM (black points). We also plot the fitted magnitude offsets for the magnitude measurements
in the range 16-18~mag (red points), and fainter than 18~mag (green points).
In each plot, the error bars represent the uncertainties in the magnitude offsets, and they are generally much smaller than
the plot symbols ($\sim$0.02-0.1~mmag). We refrain from plotting magnitude offsets
with uncertainties of greater than 5~mmag, and we mark those magnitude offsets that fall outside of the plot range with an asterisk of the
relevant colour. The vertical blue line in each plot represents the PSF FWHM corresponding to critical sampling (0.93\arcsec).
Objects in the sky scan from left to right in our plot layout. The corresponding plots for the remaining camera
columns are very similar, and for brevity, we do not reproduce them here.

We find in general that as the PSF FWHM increases, the magnitude offsets increase, indicating that the larger the PSF FWHM,
the fainter an object is measured by the SDSS pipeline aperture photometry routines with the 7.43\arcsec$\,$ fixed aperture.
We also find that this trend is much stronger for fainter objects, with the strongest trend exisiting in the $z$ waveband.
The amplitude of this trend over the PSF FWHM range is $\sim$7-15~mmag in each waveband for the magnitude measurements brighter than 16~mag,
$\sim$12-25~mmag in each waveband for the magnitude measurements in the range 16-18~mag, $\sim$30-60~mmag in the
$ugri$ wavebands for the magnitude measurements fainter than 18~mag, and $\sim$100-170~mmag in the $z$ waveband
for the magnitude measurements fainter than 18~mag. We have also tested that the form and amplitude of this trend is independent
of right ascension, declination, detector coordinates, subpixel coordinates, and the number of neighbouring
objects within 15\arcsec$\,$ by making similar plots to those displayed in
Fig.~\ref{fig:maps_ap} for various mutually-exclusive partitions of our photometric data.

The detected systematic trend in the magnitude offsets as a function of PSF FWHM
is consistent with a problem of sky-background over-subtraction in the images from which the aperture fluxes are calculated.
Sky over-subtraction by a constant amount per pixel in an image area containing
objects of a range of brightnesses implies that each of these objects will have an aperture flux measurement that is under-estimated by
a constant amount $\Delta f$, and therefore the ratio $\Delta f / f$, where $f$ is the true object aperture flux, will be
larger for the fainter objects. The result is that the fainter objects will have a larger magnitude over-estimate
$\Delta m \approx -2.5 \log_{10} (1 - \Delta f / f)$
than the brighter objects. As we have stressed in Section~\ref{sec:apermag1}, we cannot detect such a distortion of the aperture magnitude scale
itself using the repeat observations of objects of constant brightness.
However, we can detect this effect through the amplification of any PSF FWHM dependent trend 
in the magnitude offsets for fainter aperture magnitude measurements.
In our case, the amplification of the trend manifests itself as a stronger gradient in the
magnitude offsets as a function of PSF FWHM for the fainter magnitude measurements.

The fact that objects are measured as systematically fainter as the PSF FWHM increases can also easily explained by a
sky-background over-subtraction problem. An image with a larger PSF FWHM has the flux from objects spread out over more image pixels than
an image with a smaller PSF FWHM, leading to a larger fraction of supposedly pristine sky-background pixels that are contaminated by 
a small amount of flux from the PSF wings of the image objects. Inclusion of such contaminated pixels in the sky estimation is hard to avoid
due to the very low level of object flux contamination over a large area, resulting in a systematic over-estimate (and therefore over-subtraction)
of the sky background level which is manifestedly worse for images with a larger PSF FWHM.

AIH11 discuss the fact that the sky estimation procedure of the SDSS pipeline has a tendency to over-subtract the outer regions
of large galaxies on the sky as found by several authors (see references therein). We suspect that the systematic trend that we have detected here
is a symptom of the same problem for unresolved objects.

We have also considered the possibility that the detected systematic trend in the magnitude offsets
as a function of PSF FWHM is due to increased flux losses
outside of the fixed aperture when the PSF FWHM is larger. However, if this was the case, then the trend should
be magnitude independent, since the {\it fraction} of object flux lost outside of the fixed aperture is independent of the object
brightness. This is contrary to what we observe, and therefore we may be confident that this effect is not the principal cause of the
detected systematic trend, although it may still contribute to the trend. Correction of the sky-background over-subtraction
problem at the image processing stage will help to reveal whether aperture flux losses affect the aperture flux measurements as 
a function of PSF FWHM for the 7.43\arcsec$\,$ fixed aperture.

We also mention one other possible explanation for the detected systematic trend. Aperture flux losses for {\it isolated} objects are a 
function of PSF FWHM only. However, for objects that have one or more neighbouring objects whose fluxes contaminate the fixed aperture,
the amount of flux contamination, and therefore the measured aperture flux, will have a complicated dependence on PSF FWHM
(the exact form of which is determined by the spatial
distribution of the neighbouring objects relative to the aperture boundary), and the aperture magnitude measurements for fainter 
objects of interest will be affected by a larger amount by the flux contamination.
The fact that we have already found earlier in this section that the form and amplitude of the detected systematic
trend in the fitted magnitude offsets as a function of PSF FWHM is independent of the number of neighbouring objects
within 15\arcsec$\,$ allows us to rule out this explanation as a primary cause.

\subsubsection{Trends As A Function Of Subpixel Coordinates}
\label{sec:apermag3}

Now we investigate the systematic trends in the aperture magnitudes as a function of subpixel coordinates.
For each detector, we partition our data into a uniform grid of 5$\times$5 bins in subpixel coordinates covering the area of a single detector pixel
using the object centroids\footnote{We calculate the subpixel
coordinates of the object centroid as the fractional part of the pixel coordinates of the object centroid,
which are stored in the SDSS DR8 data base under the COLC and ROWC entries. The pixel
coordinates in the SDSS DR8 data base correspond to the original detector coordinates because SDSS images are not corrected
for optical distortions by resampling before object detection and photometry is performed on them (\citealt{sto2002}). This implies that our
calculated subpixel coordinates faithfully represent the subpixel coordinates of the detector pixels.},
and we further subdivide the data in each bin into those observations for which the
corresponding PSF FWHM is less than the critical sampling of 0.93\arcsec, greater than the critical sampling but less than 1.5\arcsec,
and greater than 1.5\arcsec. We construct our photometric model as in Section~\ref{sec:phot_model} adopting an unknown magnitude offset
for each bin, and we follow the iterated fitting procedure described in Section~\ref{sec:apermag1}.

In Fig.~\ref{fig:maps_ap}, above each plot of magnitude offsets versus PSF FWHM, there are three square ``images'' with
5$\times$5 elements. Each image represents the area of a single detector pixel, orientated such that objects in the sky, which scan
along the detector columns, scan from left to right. We display the fitted magnitude offsets for the observations with a PSF FWHM
that is less than the critical sampling, greater than the critical sampling but less than 1.5\arcsec, and greater than 1.5\arcsec, in the
left-hand, middle, and right-hand images, respectively. The image values are displayed in units of $\sigma$, where $\sigma$
is the uncertainty in each magnitude offset. We note that the $\sigma$ values are approximately the same
(to within $\sim$1-2\%) for each subpixel region because the photometric observations are distributed uniformly over the pixel area.
We force the mean of each image array to be zero, which is necessary for display purposes because of the general trend of the
aperture magnitudes as a function of PSF FWHM and object brightness. The colour-scale in each of the images is linear and runs between
$\mp$4$\sigma$ and $\mp$12$\sigma$ for the $ugiz$ and $r$ wavebands, respectively.

The images above each plot in Fig.~\ref{fig:maps_ap} may be interpreted as intrapixel photometric sensitivity
maps (IPSMs\footnote{An IPSM (\citealt{kav1998}) describes how the photometry of an unresolved object depends
on the position of the PSF centroid within a pixel, and it is physically influenced by the pixel response function,
the shape/width of the object PSF, and, if applicable, the TDI detector readout mode. For under-sampled images, the effect is greatest,
and as the PSF FWHM increases, the effect is washed out by the weaker PSF gradients over any one pixel.})
for the regimes of under-sampled, normally-sampled, and well-sampled PSFs (left-hand, middle, and right-hand images, respectively)
when measured by the SDSS pipeline aperture photometry routines with the 7.43\arcsec$\,$ fixed aperture.
These IPSMs show that for the $ugiz$ wavebands we do not detect any systematic trends as a function of subpixel
coordinates at the 4$\sigma$ level (although there is a hint of a systematic trend for the $i$ waveband), and they further highlight
that the scatter in the magnitude offsets is greater than the corresponding uncertainties by a factor of $\sim$1.7, which suggests
the presence of some unmodelled systematic trends.
We may translate the 4$\sigma$ levels into upper limits on the systematic trends that may exist,
and in Table~\ref{tab:dr8_data_repeat_ap}, we report these 4$\sigma$ upper limits for the regimes of under-sampled, normally-sampled, and 
well-sampled PSFs for each detector in columns 5, 6, and 7, respectively. In general, the upper limits that we derive are $\sim$0.1-0.3~mmag
for the regimes of normally-sampled and well-sampled PSFs, and $\sim$0.2-3.2~mmag for the regime of under-sampled PSFs.

\begin{figure}
\centering
\epsfig{file=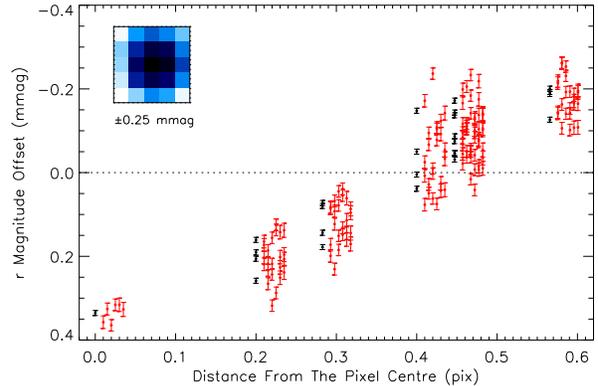,angle=0.0,width=\linewidth}
\caption{The $r$-waveband IPSMs as a function of distance from the centre of the map
         for each of the six detectors (red points). For clarity, the points in this plot corresponding to different
         detectors have been shifted along the $x$-axis by different fixed amounts. The master IPSM
         for all of the $r$ waveband data is displayed as the image in the panel in the top left-hand corner with
         a linear colour-scale between $\mp$0.25~mmag, and with the same orientation as the maps in Fig.~\ref{fig:maps_ap}.
         The corresponding radial profile of the master IPSM is plotted as the black points.
         \label{fig:radial_aper}}
\end{figure}

For the $r$ waveband, the IPSMs in Fig.~\ref{fig:maps_ap} reveal a smooth,
approximately radially-symmetric, systematic trend detected with a $\sim$20$\sigma$ range between the extreme values.
The trend appears to be present for all six detectors in each PSF sampling regime.
Hence, we have recalculated the IPSMs for each detector in the $r$ waveband including
all data regardless of the PSF FWHM. Then, in Fig.~\ref{fig:radial_aper}, for each detector, we plot the magnitude
offsets from these IPSMs as a function of distance from the centre of the map (red points).
Note that for clarity we have shifted the points corresponding to different detectors by different
fixed amounts along the $x$-axis. Clearly the systematic trends are very similar between the six detectors, and therefore we
have also derived a single master IPSM for all of the $r$ waveband data, and we present this map
in Fig.~\ref{fig:radial_aper} as the image panel in the top left-hand corner. We also plot the radial profile of this master
IPSM as the black points. Finally, we note that the form and amplitude of this
IPSM is independent of right ascension, declination, detector coordinates, object brightness, and the number
of neighbouring objects within 15\arcsec.

The peak-to-peak amplitude of the $r$-waveband master IPSM is $\sim$0.54~mmag,
and it shows that objects observed with their centroid at the centre of a detector pixel are
measured as $\sim$0.54~mmag fainter than if their centroid is at the corner.
This is in contrast to what we would expect if this systematic trend were due to the pixel response
function (PRF), which generally results in objects being measured as {\it brighter} when their centroid is in the centre
of a pixel compared to when their centroid is at the corner/edge (e.g. \citealt{kav1998}).
We also find that the detected systematic
trend as a function of subpixel coordinates is {\it independent} of the PSF FWHM, and therefore it cannot be
the result of the PRF convolved with the object PSF. 
Furthermore, we do not observe the asymmetry we would expect in the IPSM
for detectors using the TDI readout mode (\citealt{gib1992}). This
leads us to believe that the systematic trend is an effect introduced by the algorithms used to measure the aperture magnitudes, which is not
so surprising when we find in Section~\ref{sec:psfmag1} that the SDSS pipeline introduces a variety of systematic trends
into the PSF magnitudes.

\subsubsection{Correcting The Aperture Magnitudes}
\label{sec:apmagcorr}

\begin{figure}
\centering
\epsfig{file=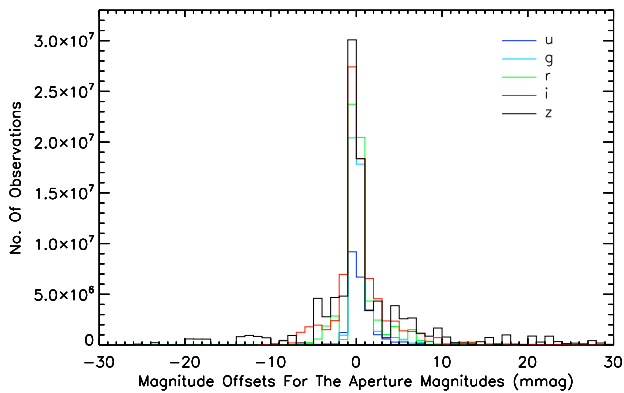,angle=0.0,width=\linewidth}
\caption{Histograms of the magnitude offsets (mmag) for each filter that we calculate for use in
         calibrating the aperture magnitude measurements for the full set of selected observations from SDSS DR8
         in Table~\ref{tab:dr8_data}, using data for the six camera columns combined.
         \label{fig:hist_aper_corr}}   
\end{figure}

Ideally, the systematic photometric trends that we have detected in the aperture magnitudes
could be corrected by improving the image processing algorithms in the SDSS
pipeline and then reprocessing the imaging data. However, this solution is outside of the scope of this paper.
Instead, we opt to correct the SDSS DR8 aperture magnitudes by subtracting the
fitted magnitude offsets $Z_{k}$ from the aperture magnitude measurements, which constitutes a post-processing
self-calibration of the data. Since the correction of SDSS DR8 aperture magnitudes is of general interest to the astronomical community,
we supply an {\tt IDL} program (...weblink...) which can be used to correct a set of aperture magnitude measurements from SDSS DR8 for the systematic
trends that we detected as a function of PSF FWHM and object brightness in Section~\ref{sec:apermag2}. We do not include the corrections for the $r$-waveband
master IPSM because of the very small amplitude of this effect. Note that these
magnitude corrections should only be applied to aperture magnitude measurements of point sources (since they were derived exclusively from such sources), and we
stress that even though the subtraction of our fitted
magnitude offsets from the aperture magnitudes constitutes a {\it relative} photometric calibration, any {\it absolute}
calibration of the photometry may need to be redetermined.

We may now assess the significance of our photometric calibrations. We do this by first
fitting the photometric model in Equation~\ref{eqn:phot_model} with no magnitude offset parameters to
the aperture magnitude measurements in the full set of repeat observations in Table~\ref{tab:dr8_data_repeat_ap}, and
we calculate the associated chi-squared $\chi^{2}_{\mbox{\scriptsize before}}$. We then subtract the magnitude offsets $Z_{k}$
corresponding to the detected systematic trends as a function of PSF FWHM and object brightness in Section~\ref{sec:apermag2}, and which have an uncertainty
of less than 5~mmag, from the appropriate aperture magnitude measurements. Finally,
we fit again the photometric model in Equation~\ref{eqn:phot_model} with no magnitude offset
parameters to the corrected aperture magnitude measurements and calculate the associated chi-squared $\chi^{2}_{\mbox{\scriptsize after}}$.

In Table~\ref{tab:dr8_data_repeat_ap}, we tabulate the values of 
$\Delta \chi^{2} = \chi^{2}_{\mbox{\scriptsize before}} - \chi^{2}_{\mbox{\scriptsize after}}$. By fitting
and applying corrections to the aperture magnitude measurements, we have introduced $N_{\mbox{\scriptsize z}} \sim 90$ new parameters
between calculating the different $\chi^{2}$ values, and we would therefore expect that $\Delta \chi^{2} \sim 90$ if there are no systematic trends.
The values of $\Delta \chi^{2}$ tabulated in Table~\ref{tab:dr8_data_repeat_ap} are much greater than $N_{\mbox{\scriptsize z}}$
which indicates that our corrections to the repeat aperture magnitude measurements are highly significant. We note that the variance of $\chi^{2}$
is $2 N_{\mbox{\scriptsize dof}}$, where $N_{\mbox{\scriptsize dof}}$ is the number of degrees of freedom. Hence, the significance in units of $\sigma$ of
the $\Delta \chi^{2}$ values is given by $\Delta \chi^{2} / \sqrt{2 N_{\mbox{\scriptsize dof}}}$.
For $\chi^{2}_{\mbox{\scriptsize before}}$, we have
$N_{\mbox{\scriptsize dof}} = N_{\mbox{\scriptsize data}} - N_{\mbox{\scriptsize obj}}$ (column 4 minus column 3 of Table~\ref{tab:dr8_data_repeat_ap}),
and for $\chi^{2}_{\mbox{\scriptsize after}}$, we have
$N_{\mbox{\scriptsize dof}} = N_{\mbox{\scriptsize data}} - N_{\mbox{\scriptsize obj}} - N_{\mbox{\scriptsize z}} \approx 
N_{\mbox{\scriptsize data}} - N_{\mbox{\scriptsize obj}}$
(since $N_{\mbox{\scriptsize z}} \ll N_{\mbox{\scriptsize obj}}$).
In column 9 of Table~\ref{tab:dr8_data_repeat_ap}, we list the quantity $\Delta \chi^{2} / \sqrt{2 ( N_{\mbox{\scriptsize data}} - N_{\mbox{\scriptsize obj}} )}$
for each detector. We conclude that our corrections to the repeat aperture magnitudes are significant at the
$\sim$20-40$\sigma$, 58-109$\sigma$, 74-120$\sigma$, 56-84$\sigma$, and 45-68$\sigma$ level
for the $ugriz$ wavebands, respectively. We also note that if we reproduce Fig.~\ref{fig:maps_ap} using the corrected aperture magnitude measurements, then the systematic
trends as a function of PSF FWHM are not present any more, which confirms that applying the corrections has successfully calibrated the aperture magnitudes
to compensate for this particular systematic trend.

To conclude with our analysis of the SDSS aperture magnitudes, in Fig.~\ref{fig:hist_aper_corr} we plot for each filter a histogram of the magnitude offsets (mmag)
that we calculate for use in calibrating
the aperture magnitude measurements for the full set of selected observations from SDSS DR8 in Table~\ref{tab:dr8_data}, combining the data for the six camera columns. We see that
there are a significant number of corrections that have absolute values $>$2~mmag and stretching up to $\sim$10~mmag and beyond, where $\sim$10~mmag is the point at which the
distribution seems to flatten out in both directions (there are also some magnitude corrections that lie outside of the plot limits).
These magnitude corrections are of the order of the quoted
$\sim$1\% precision of the SDSS DR8 photometric measurements, and we believe that the application of these corrections to the aperture magnitude measurements will significantly improve the SDSS DR8
photometric precision in all filters.

\begin{table}
\caption{The properties of the selected repeat observations with PSF magnitude measurements from SDSS DR8 (see Section~\ref{sec:repeat}) for each
         of the 30 SDSS imaging detectors. The data sets for each detector are organised by filter and camera column. In column 5, we report the
         amount by which the $\chi^{2}$ for the repeat observations decreases after calibrating the PSF magnitudes using the fitted magnitude offsets from
         Sections~\ref{sec:psfmag2}~\&~\ref{sec:psfmag3}. In column 6, we list the $\Delta \chi^{2}$ values in units of $\sigma$.
         }
\centering
\begin{tabular}{@{}cccccc@{}}
\hline
Filter & Camera & No. Of  & No. Of       & $\Delta \chi^{2}$ & $\Delta \chi^{2} / \sqrt{2 ( N_{\mbox{\scriptsize data}} - N_{\mbox{\scriptsize obj}} )}$ \\
       & Column & Objects & Observations &                   &       \\
\hline
$u$    & 1      &  568101 & 1695234      &  21266            &  14.2 \\
$u$    & 2      &  630323 & 1883568      &  21131            &  13.3 \\
$u$    & 3      &  624662 & 1874595      &  32228            &  20.4 \\
$u$    & 4      &  639716 & 1916424      &  20993            &  13.1 \\
$u$    & 5      &  624109 & 1869581      &  17378            &  11.0 \\
\smallskip   
$u$    & 6      &  651562 & 1957703      &  12988            &   8.0 \\
$g$    & 1      & 1269177 & 3704174      & 109859            &  49.8 \\
$g$    & 2      & 1276445 & 3739691      &  76509            &  34.5 \\
$g$    & 3      & 1258048 & 3690359      & 291155            & 132.0 \\
$g$    & 4      & 1255473 & 3680556      & 129834            &  59.0 \\
$g$    & 5      & 1308813 & 3845975      & 125528            &  55.7 \\
\smallskip
$g$    & 6      & 1302986 & 3833631      &  92512            &  41.1 \\
$r$    & 1      & 1879663 & 5435094      & 203461            &  76.3 \\
$r$    & 2      & 1862306 & 5396597      &  84934            &  31.9 \\
$r$    & 3      & 1877275 & 5445389      & 169107            &  63.3 \\
$r$    & 4      & 1887768 & 5478996      & 181105            &  67.6 \\
$r$    & 5      & 1807342 & 5256469      &  83999            &  32.0 \\
\smallskip
$r$    & 6      & 1966558 & 5741314      & 121363            &  44.2 \\
$i$    & 1      & 2379513 & 6899266      & 102726            &  34.2 \\
$i$    & 2      & 2523293 & 7330526      & 128375            &  41.4 \\
$i$    & 3      & 2318811 & 6729380      &  82222            &  27.7 \\
$i$    & 4      & 2386082 & 6927950      & 206118            &  68.4 \\
$i$    & 5      & 2607420 & 7606943      & 124378            &  39.3 \\
\smallskip
$i$    & 6      & 2499355 & 7302803      & 154839            &  50.0 \\
$z$    & 1      & 3133930 & 9112708      &  37810            &  10.9 \\
$z$    & 2      & 3190973 & 9285784      &  73341            &  21.0 \\
$z$    & 3      & 3212432 & 9363787      &  41056            &  11.7 \\
$z$    & 4      & 3255562 & 9482727      &  31237            &   8.9 \\
$z$    & 5      & 3115876 & 9097166      &  32328            &   9.3 \\
$z$    & 6      & 3007108 & 8787324      &  39494            &  11.6 \\
\hline
\end{tabular}
\label{tab:dr8_data_repeat_psf}
\end{table}

\subsection{PSF Magnitudes}
\label{sec:psfmag1}

We now analyse the PSF magnitudes. We use the data corresponding to Table~\ref{tab:dr8_data_repeat_psf}.
Independently for each detector, we investigate how the PSF magnitudes correlate with various object/image properties.
Using the same fitting procedure as that described in Section~\ref{sec:apermag1}
applied to the PSF magnitude measurements, we find that there are no significant trends in the fitted magnitude offsets as a function
of detector row ($Z_{k} < 0.2$~mmag) or column ($Z_{k} < 0.5$~mmag), and, as with the aperture magnitude measurements,
we do find a systematic trend in the magnitude offsets as a function of PSF FWHM and object brightness.
We also find a systematic trend in the magnitude offsets as a function of subpixel coordinates.

\begin{figure*}
\centering
\epsfig{file=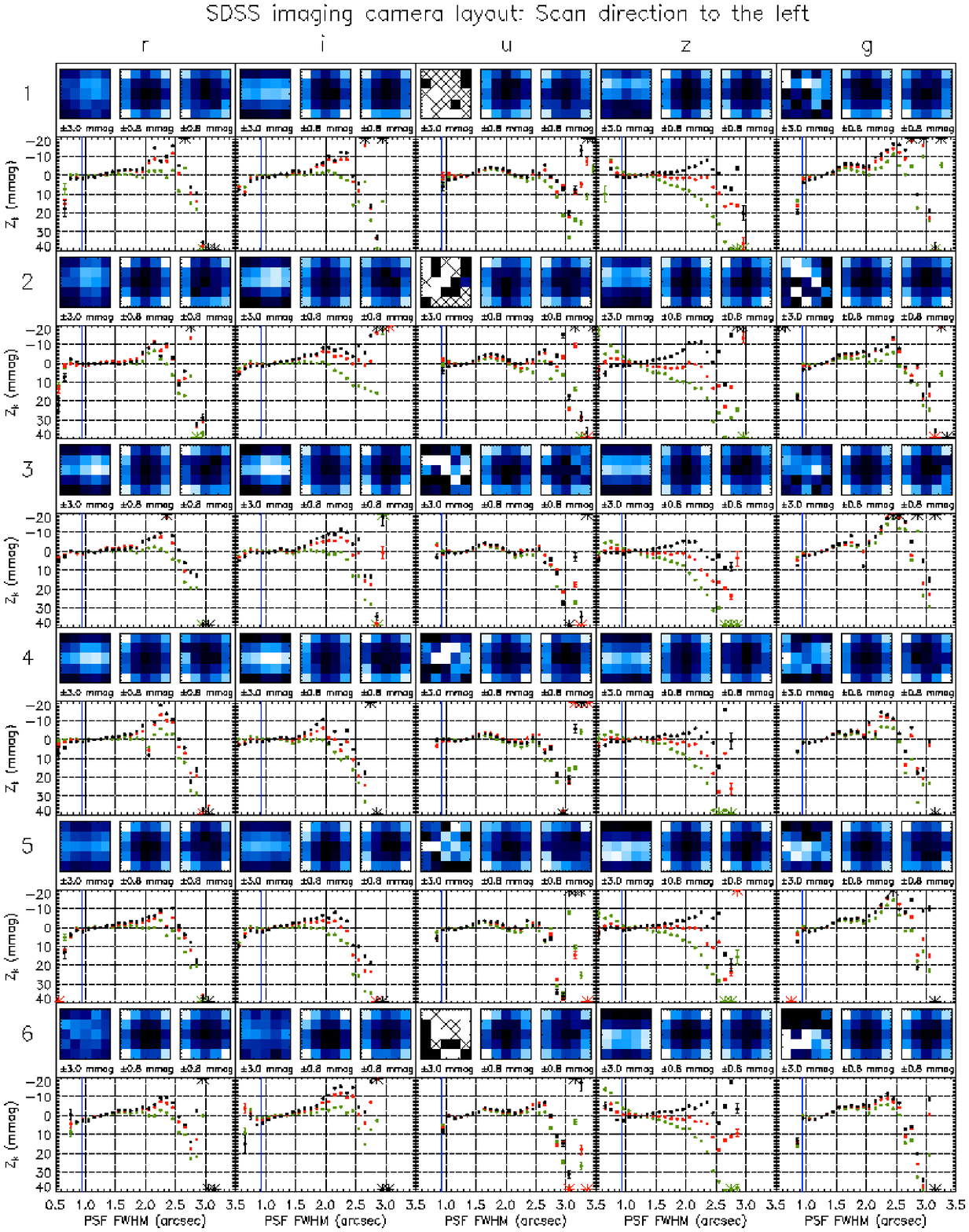,angle=0.0,width=0.99\linewidth}
\caption{Systematic trends in the fitted magnitude offsets $Z_{k}$ for the PSF magnitude measurements
         as a function of PSF FWHM, object brightness, and subpixel coordinates
         (see text in Sections~\ref{sec:psfmag2}~\&~\ref{sec:psfmag3} for details).
         The data used to generate these plots are the selected repeat observations
         with PSF magnitude measurements from SDSS DR8 (see Section~\ref{sec:repeat}).
         The black, red, and green points correspond to PSF magnitude measurements
         brighter than 16~mag, in the range 16-18~mag, and fainter than 18~mag, respectively.
         Asterisks represent magnitude offsets that fall outside of the plot range.
         The three images above each plot represent the IPSMs
         for the regimes of under-sampled, normally-sampled, and well-sampled PSFs (from left to right).
         The colour-scale is linear and runs between $\mp$3~mmag for the under-sampled PSF regime,
         and between $\mp$0.8~mmag for the normally-sampled and well-sampled PSF regimes.
         \label{fig:maps_psf}}
\end{figure*}

\subsubsection{Trends As A Function Of PSF FWHM}
\label{sec:psfmag2}

Fig.~\ref{fig:maps_psf} is the equivalent of Fig.~\ref{fig:maps_ap} produced for the PSF magnitudes.
The only difference in the plot schema is that in Fig.~\ref{fig:maps_psf}, the colour-scale in each of the IPSMs
is in units of mmag rather than in units of $\sigma$. For all wavebands, the colour-scale
in these maps is linear, and it runs between $\mp$3~mmag for the maps corresponding to the under-sampled PSF
regime, and between $\mp$0.8~mmag for the maps corresponding to the normally-sampled and well-sampled PSF regimes.

Inspection of the plots of the fitted magnitude offsets as a function of PSF FWHM reveals that
as the PSF FWHM increases from the critical sampling to $\sim$2.2-2.5\arcsec, the magnitude offsets ``oscillate''
with an amplitude of $\sim$2-6~mmag (the $ug$ wavebands show the clearest examples) while also decreasing with a slight gradient
($\sim$1-5~mmag/arcsec; most notably in the $g$ waveband). This behaviour is largely independent of object brightness (except in the $z$ waveband).
It is also interesting to note that the form and amplitude of the trend is consistent between the different camera columns
for each waveband. For PSF FWHMs greater than $\sim$2.2-2.5\arcsec, we find that as the PSF FWHM increases,
the magnitude offsets increase sharply by up to $\sim$20-50~mmag. For PSF FWHMs less than the
critical sampling, the systematic trend tends to be that as the PSF FWHM decreases, the fainter an object is measured (except in the $z$ waveband), with
some evidence that this behaviour becomes more pronounced for even smaller PSF FWHMs (see the $ri$ wavebands).

We have tested that the detected systematic trends in the magnitude offsets as a function of PSF FWHM
are independent of right ascension, declination, detector coordinates, subpixel coordinates, and the number of
neighbouring objects within 15\arcsec$\,$ by making similar plots to those displayed in Fig.~\ref{fig:maps_psf}
for various mutually-exclusive partitions of our photometric data.

Clearly, the systematic trends in the magnitude offsets for the PSF magnitude measurements are more complicated than the trends observed in the 
magnitude offsets for the aperture magnitude measurements,
and this suggests that it is the algorithms used to measure the PSF magnitudes that introduce these trends into the photometry.
The scientific literature contains only brief qualitative descriptions of the PSF fitting procedures implemented in the SDSS pipeline
(\citealt{lup2001}; \citealt{sto2002}), and there are no published studies in which an analysis of the behaviour of these routines has
been performed. Therefore, we do not attempt to offer an in-depth explanation for any of these systematic trends in terms 
of the algorithms used to measure the PSF magnitudes. We limit our scope to simply highlighting the existence and form of the systematic trends in the
PSF magnitude measurements, and to speculating that the ``oscillations'' present in the magnitude offsets
are probably due to some form of resonance between the PSF shape and the underlying pixel grid.

\subsubsection{Trends As A Function Of Subpixel Coordinates}
\label{sec:psfmag3}

\begin{figure}
\centering
\epsfig{file=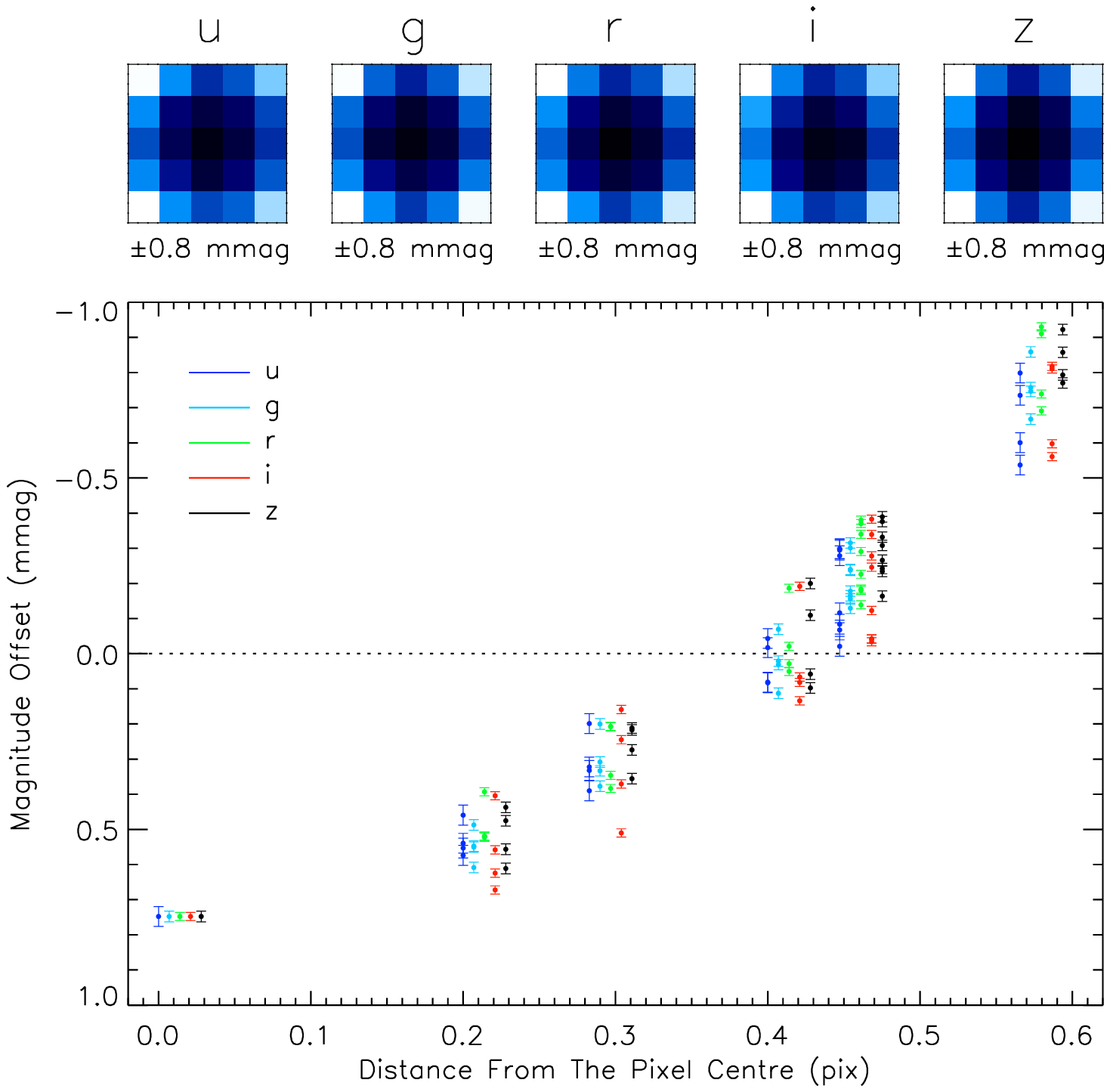,angle=0.0,width=\linewidth}
\caption{The master IPSMs for each waveband as a function of distance from the centre of the map.
         These master IPSMs have been derived for each waveband using all of the PSF magnitude measurements over the six detectors
         that have corresponding PSF FWHMs greater than the critical sampling.
         For clarity, the points in this plot corresponding to different wavebands have been plotted with different colours and shifted
         along the $x$-axis by different fixed amounts. The $u$-waveband IPSM has a mean value of zero, and 
         for comparison purposes, the maps for the other wavebands have had a constant added to the magnitude offsets
         so as to match the central magnitude offset between the maps. The images in the top panels
         display the master IPSMs for each waveband as a function of subpixel coordinates using a linear colour-scale
         between $\mp$0.8~mmag, and with the same orientation as the maps in Fig.~\ref{fig:maps_psf}.
         \label{fig:radial_psf}}
\end{figure}

The IPSMs displayed in Fig.~\ref{fig:maps_psf} show very clear systematic trends in each of the
regimes of under-sampled, normally-sampled, and well-sampled PSFs. For the regime of under-sampled PSFs, we find that in the $riz$
wavebands, objects with their centroid close to the centre of a detector pixel are measured by up to $\sim$4-7~mmag brighter
than if their centroid is at the corner of a detector pixel. We cannot discern this effect for the regime of under-sampled
PSFs in the $ug$ wavebands due to the lack of data in these wavebands with PSF FWHMs less than the critical sampling. Since these
systematic trends are not present in the regime of under-sampled PSFs for the aperture magnitude measurements, we can only conclude
that they are {\it introduced} into the PSF magnitude measurements by the algorithms used to measure the PSF magnitudes,
and that the underlying cause is not a physical effect (i.e. a non-uniform PRF convolved with the object PSF).
We speculate that a possible explanation for the trends could be a slight mismatch between the PSF model and the actual shape of the PSF.
The fact that, for the regime of under-sampled PSFs, the exact form of the IPSMs varies between different wavebands and camera
columns hints that the deficiencies in the PSF modelling depend on the exact PSF shape, which is unique to each detector.

For PSF magnitude measurements with PSF FWHMs that are greater than the critical sampling, we find that for all wavebands, the 
IPSMs in Fig.~\ref{fig:maps_psf} have the same smooth and approximately radially-symmetric form
as the systematic trend detected for the aperture magnitudes in the $r$ waveband in Section~\ref{sec:apermag3}.
For each waveband, the trends are very similar in form and amplitude between the six detectors and the normally-sampled
and well-sampled PSF regimes. Therefore we have derived a single master IPSM
for each waveband using all of the PSF magnitude measurements over the six detectors that have corresponding PSF FWHMs greater
than the critical sampling. We present these maps in Fig.~\ref{fig:radial_psf} as the image panels along the top of the figure.
In the same figure, we also plot the radial profiles of the master IPSMs for each waveband.
Finally, we note that the form and amplitude of these IPSMs are independent of right ascension, declination,
detector coordinates, object brightness, and the number of neighbouring objects within 15\arcsec.

The master IPSMs displayed in Fig.~\ref{fig:radial_psf} are very similar between wavebands
with peak-to-peak amplitudes of $\sim$1.55, 1.61, 1.68, 1.57, and 1.67~mmag for the $ugriz$ wavebands, respectively. These amplitudes
are approximately three times larger than the peak-to-peak amplitude of the 
same systematic trend detected for the $r$-waveband aperture magnitude measurements. Again, we believe that this
systematic trend is an effect introduced by the algorithms used to measure the PSF magnitudes.

\subsubsection{Correcting The PSF Magnitudes}
\label{sec:psfmagcorr}

\begin{figure}
\centering
\epsfig{file=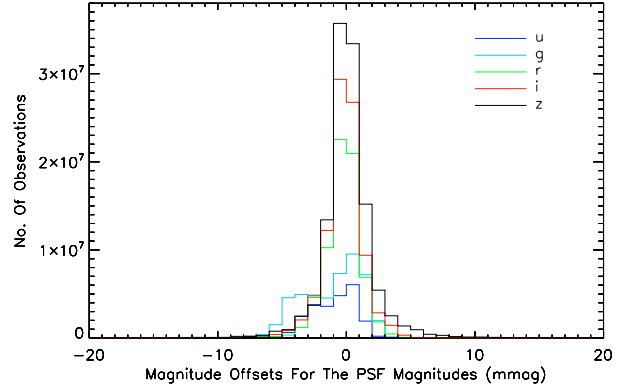,angle=0.0,width=\linewidth}
\caption{Histograms of the magnitude offsets (mmag) for each filter that we calculate for use in
         calibrating the PSF magnitude measurements for the full set of selected observations from SDSS DR8
         in Table~\ref{tab:dr8_data}, using data for the six camera columns combined.
         \label{fig:hist_psf_corr}}
\end{figure}

Our {\tt IDL} program described in Section~\ref{sec:apmagcorr} (...weblink...) may also be used to correct
a set of PSF magnitude measurements of point sources from SDSS DR8. The program subtracts the appropriate fitted magnitude offsets $Z_{k}$
corresponding to the systematic trends that we detected as a function of PSF FWHM and object brightness in Section~\ref{sec:psfmag2}
from the PSF magnitude measurements to be corrected. Also, for each detector, we repeat the analysis from Section~\ref{sec:psfmag3} to
derive two IPSMs, one for each of the regimes of PSF FWHMs less than and greater than the critical
sampling. For each map, we force the mean of the magnitude offsets to be zero to minimise the effect on the absolute photometric
calibration when they are applied. The program subtracts the appropriate magnitude offsets taken from these IPSMs
from the PSF magnitude measurements to correct for the systematic trends that we detected as a function of subpixel coordinates in
Section~\ref{sec:psfmag3}. For the $ug$ wavebands, we refrain from applying the magnitude offsets from the IPSMs
calculated for the regime of under-sampled PSFs because these maps are very noisy due to the lack of
appropriate data.

In essence, the procedure for calibrating the PSF magnitudes is the same as for the aperture magnitudes
except for the inclusion of the extra correction for the systematic trends as a function of subpixel coordinates, which
introduces a further 50 parameters into the model for the photometric data. However, the number of extra parameters introduced for our
corrections to the PSF magnitude measurements is still insignificant compared to the total number of free parameters. Therefore, we follow
a similar analysis to that of the aperture magnitudes with regards to the significance of our corrections
to the repeat PSF magnitude measurements, and in columns 5 and 6 of Table~\ref{tab:dr8_data_repeat_psf}, we tabulate the values of
$\Delta \chi^{2}$ and $\Delta \chi^{2} / \sqrt{2 ( N_{\mbox{\scriptsize data}} - N_{\mbox{\scriptsize obj}} )}$
for each detector. We conclude that our corrections to the repeat PSF magnitude measurements are significant at the 
$\sim$8-20$\sigma$, 35-132$\sigma$, 32-76$\sigma$, 28-68$\sigma$, and 9-21$\sigma$ level
for the $ugriz$ wavebands, respectively. We also note that if we reproduce Fig.~\ref{fig:maps_psf} using the corrected PSF magnitude measurements,
then the systematic trends as a function of PSF FWHM and subpixel coordinates are not present any more, which confirms that applying
the corrections has successfully calibrated the PSF magnitudes to compensate for these particular systematic trends.

In Fig.~\ref{fig:hist_psf_corr} we plot for each filter a histogram of the magnitude offsets (mmag) that we calculate for use in calibrating
the PSF magnitude measurements for the full set of selected observations from SDSS DR8 in Table~\ref{tab:dr8_data}, combining the data for the six camera columns.
Again, we see that there are a significant number of corrections that have absolute values $>$2~mmag and stretching up to $\sim$8~mmag.
As with the aperture magnitudes, these corrections are of the order of the quoted
$\sim$1\% precision of the SDSS DR8 photometric measurements, and we believe that the application of these corrections to the PSF magnitude measurements will significantly 
improve the SDSS DR8 photometric precision in all filters.

\section{Comparing The SDSS Aperture And PSF Magnitude Scales}
\label{sec:compare}

\begin{figure*}
\centering
\epsfig{file=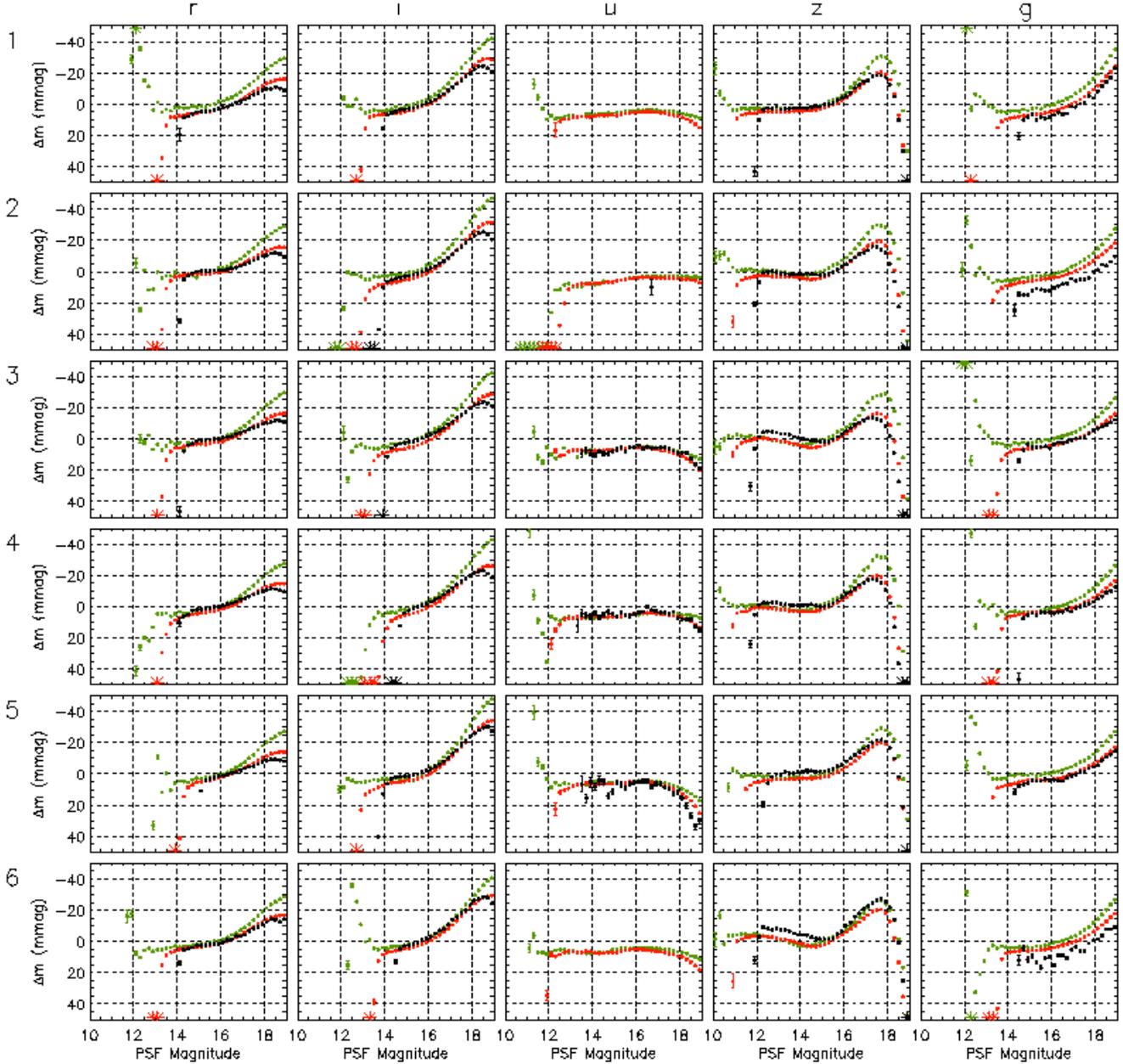,angle=0.0,width=\linewidth}
\caption{The mean difference between the PSF and aperture magnitude measurements as a function of object brightness (PSF magnitude).
         The data used to generate these plots are the selected observations from SDSS DR8 restricted to observations
         of objects that have no neighbouring objects within 15\arcsec. The black, red, and green points correspond to the regimes of
         under-sampled, normally-sampled, and well-sampled PSFs, respectively. Asterisks represent mean differences that fall outside
         of the plot range.
         \label{fig:scale_comp}}
\end{figure*}

In this section, we investigate how the SDSS PSF magnitude scale relates to the SDSS aperture magnitude scale for the
7.43\arcsec$\,$ fixed aperture. This investigation is important because the photometric calibration parameters derived by
PAD08 are obtained by fitting the aperture magnitudes and the calibration is subsequently applied to the PSF magnitudes.
This is a valid procedure for the calibration of the PSF magnitudes as long as they follow
the same magnitude scale (to within the addition of an arbitrary constant) as the aperture magnitudes. Any deviation
of the PSF magnitude scale from the aperture magnitude scale other than a simple offset renders the calibration
parameters derived from the aperture magnitudes as incompatible with the PSF magnitudes. Consequently, in such a case,
the application of the calibration parameters to the PSF magnitudes will introduce undesirable systematic errors.

We may use the selected observations from SDSS DR8 in Table~\ref{tab:dr8_data} to compare the PSF magnitude measurements to the
aperture magnitude measurements. We analyse the photometric data for each detector independently. Firstly,
we drop the observations for which the SDSS pipeline failed to measure an aperture flux. In addition, we restrict
our sample of observations to those objects which have no neighbouring objects within 15\arcsec. We do this to avoid the
inclusion of objects with aperture magnitude measurements that have been contaminated by the flux of nearby objects, which
would lead to artificially bright aperture magnitude measurements for these objects when compared to the corresponding PSF magnitude measurements.
Next, for each observation, we calculate the magnitude difference $\Delta m = m_{\mbox{\scriptsize psf}} - m_{\mbox{\scriptsize ap}}$, where
$m_{\mbox{\scriptsize ap}}$ and $m_{\mbox{\scriptsize psf}}$ are the measured aperture and PSF magnitudes, respectively,
calibrated using the procedures described in Sections~\ref{sec:apmagcorr}~\&~\ref{sec:psfmagcorr}. The uncertainty on
$\Delta m$ is calculated as the square root of the sum in quadrature of the corresponding aperture and PSF magnitude measurement
uncertainties.

We partition the $\Delta m$ values into three groups based on their corresponding PSF FWHMs; namely, the under-sampled PSFs with
PSF FWHMs less than the critical sampling of 0.93\arcsec, the normally-sampled PSFs with PSF FWHMs greater than the critical sampling
but less than 1.5\arcsec, and the well-sampled PSFs with PSF FWHMs greater than 1.5\arcsec. Then, for each grouping, we assign
the $\Delta m$ values to magnitude bins of width 0.2~mag using their corresponding PSF magnitudes, and
we calculate the inverse-variance weighted mean of the $\Delta m$ values in each bin. In Fig.~\ref{fig:scale_comp}, for each detector,
we plot these mean $\Delta m$ values as a function of PSF magnitude using black, red, and green points for the
regimes of under-sampled, normally-sampled, and well-sampled PSFs, respectively. In each plot, the error bars represent the uncertainties
in the mean $\Delta m$ values, and they are generally much smaller than the plot symbols ($\sim$0.03-0.2~mmag). We refrain
from plotting mean $\Delta m$ values with uncertainties of greater than 5~mmag, and we mark mean $\Delta m$ values that fall
outside of the plot range with an asterisk of the relevant colour. The plot panels are arranged in the same
grid layout as in Fig.~\ref{fig:maps_psf}.

If the PSF magnitude scale is the same as the aperture magnitude scale except for a difference in photometric zero-points, then, independent of object brightness, the
expected value of $\Delta m$ is the zero-point difference, and the plots of the mean $\Delta m$ values as a function of object brightness will be horizontal straight lines with a $y$-axis
value equal to the zero-point difference. However, the fact that {\it all} of the plots in Fig.~\ref{fig:scale_comp} show mean $\Delta m$ values
that depend on the object brightness in a non-linear fashion (i.e. they do not form horizontal or inclined straight lines) indicates
that the PSF magnitude scale is related to the aperture magnitude scale by a non-linear transformation. As a further complication,
we may also see in Fig.~\ref{fig:scale_comp} that this non-linear transformation is PSF FWHM dependent since the mean $\Delta m$ values
show different behaviour for each of the regimes of under-sampled, normally-sampled, and well-sampled PSFs.

We observe that, for any single waveband and PSF FWHM regime combination, the form and amplitude of the non-linear
relation between the aperture and PSF magnitude scales is very similar for all six detectors. For the $griz$ wavebands,
the fainter an object is, the brighter the PSF magnitude measurement is compared to the aperture magnitude measurement,
where the difference between the two types of magnitude measurements can reach up to $\sim$15-35~mmag for the faintest objects in our data set when
observed under normal-seeing conditions. For the $u$ waveband, we detect the opposite effect (except for camera column 2) with a difference of
up to $\sim$10-25~mmag between the aperture and PSF magnitudes for the faintest objects in our data set when observed under normal-seeing conditions.
We also note that for the $u$ waveband, the PSF magnitudes are fainter than the aperture magnitudes by $\sim$5-10~mmag, indicating that
there is a zero-point difference between the two magnitude scales for this waveband. For the $z$ waveband, the aperture and PSF
magnitude scales diverge rapidly for objects with brightnesses in the range 18-19~mag. Also, for all wavebands, we find that the
amplitude of the non-linearity in the relation between the aperture and PSF magnitude scales is greater for larger PSF FWHMs.
Finally, we note that the erratic behaviour of the mean $\Delta m$ values at the bright-end for all wavebands is most likely a consequence
of image saturation for the corresponding observations.

We have tested that the detected non-linear relation between the aperture and PSF magnitude scales is independent of right ascension, declination,
detector coordinates, and subpixel coordinates by making similar plots to those displayed in Fig.~\ref{fig:scale_comp} for various
mutually-exclusive partitions of our photometric data. Furthermore, the plots in Fig.~\ref{fig:scale_comp} are only marginally affected if we
do not calibrate the aperture and PSF magnitudes using the procedures described in Sections~\ref{sec:apmagcorr}~\&~\ref{sec:psfmagcorr}.
This is to be expected since, by necessity, our magnitude corrections are calculated independently of object brightness.

It is important to understand that our analysis may only determine how the PSF magnitude scale relates to the aperture magnitude scale. Due to the
fact that there is a non-linear relation between the aperture and PSF magnitude scales, we may deduce that {\it at least one} of the scales must deviate
from a pure asinh-magnitude scale. However, it is quite possible that {\it both} the aperture and PSF magnitude scales actually deviate from
a pure asinh-magnitude scale, and that the non-linear relation that we find between the magnitude scales simply indicates that they differ from each
other in a non-linear fashion. One way to investigate if a magnitude scale actually deviates from a pure asinh-magnitude scale is to simulate a set
of images containing objects of known magnitudes, and then to process these images with the SDSS pipeline to extract the relevant magnitude measurements
for comparison with the known magnitudes. We note that both the aperture and PSF magnitude measurements that we have employed in this analysis
have already been calibrated using the PAD08 modelling scheme. However, the calibrations will not have affected the relation between the aperture and
PSF magnitude scales because, for each observation, the same calibration correction is applied to each type of magnitude measurement.

We find that we can explain the main features of the non-linear relation between the aperture and PSF magnitude scales as, yet again, a consequence of
sky over-subtraction in the SDSS images. The SDSS pipeline performs PSF photometry by scaling the image PSF to each object on the sky-subtracted images
without fitting a local background (R.~Lupton, private communication). Hence, the main difference between the SDSS aperture and PSF photometry
is the pixel weighting scheme that is used when measuring the flux of an object. Aperture photometry weights all pixels equally within the aperture,
whereas PSF photometry weights the pixels in the aperture using the image PSF, giving more weight to the image pixels that contain most of the object
flux. Hence, the effective number of pixels used in the PSF photometry measurement is less than the effective number of pixels used in the aperture
photometry measurement, and so we would expect that if the sky background is over-subtracted from the SDSS images, then the PSF photometry measurement
will be brighter than the aperture photometry measurement. Also, we have already seen in Section~\ref{sec:apermag2} that the aperture magnitude
measurements of fainter objects suffer from larger systematic errors due to incorrect sky-subtraction, and so we would expect the PSF magnitudes
to differ by a greater amount from the aperture magnitudes for fainter objects. These effects constitute the main features of the non-linear relation
that we detect in Fig.~\ref{fig:scale_comp} between the aperture and PSF magnitude scales.

We note that the sky over-subtraction hypothesis cannot explain the PSF FWHM dependence of the relation between the aperture and PSF magnitude scales.
As the PSF FWHM increases, the PSF photometry pixel-weighting scheme tends towards the aperture photometry pixel-weighting scheme, and therefore,
in the case of sky over-subtraction, we would expect the difference between the aperture and PSF magnitude measurements to decrease, which
is the opposite of what we detect.

\section{Summary And Discussion}
\label{sec:disc}

We have investigated the systematic trends in the SDSS DR8 photometric data, extending the internal consistency checks performed
by PAD08 on the aperture magnitude residuals from their photometric calibration model. From our analysis of the repeat observations in
SDSS DR8, we have discovered a systematic trend in the aperture magnitudes that is a function of PSF FWHM, the amplitude of
which increases for fainter objects. This trend is present at the $\sim$7-15~mmag and $\sim$30-170~mmag level for the brightest
(near the saturation limit) and faintest (19~mag) objects that we analysed, respectively, with the $z$ waveband exhibiting the strongest trend.
We have also detected a low-amplitude ($\sim$0.54~mmag) systematic trend in the $r$-waveband aperture magnitudes that is a function
of subpixel coordinates.

For the SDSS PSF magnitudes, we have discovered complicated systematic trends that are a function of PSF FWHM, object brightness,
and subpixel coordinates. The trends that are a function of PSF FWHM and object brightness are present at the $\sim$10-20~mmag 
and $\sim$20-50~mmag level for PSF FWHMs that are less than $\sim$2.5\arcsec$\,$ and greater than $\sim$2.5\arcsec, respectively.
Again, it is the $z$-waveband PSF magnitudes that exhibit the strongest trends as a function of PSF FWHM and object brightness.
The systematic trends in the PSF magnitudes that are a function of subpixel coordinates are clearly detected in all wavebands
at the $\sim$1.6~mmag level for PSF FWHMs greater than the critical sampling (0.93\arcsec), and in the $riz$ wavebands at the $\sim$4-7~mmag
level for PSF FWHMs less than the critical sampling. In the $ug$ wavebands, we cannot discern these trends as a function of subpixel coordinates for the
regime of under-sampled PSFs due to the lack of appropriate data.

To address the problem of these systematic trends, we have described a method for self-calibration of the SDSS photometric data which successfully
removes the trends, and we provide an {\tt IDL} program which can be used to calibrate a set of aperture and/or PSF magnitude measurements of point sources from SDSS DR8. However,
it would be better to eliminate these trends by identifying the algorithm(s) in the SDSS pipeline that introduce them, and then improving the relevant
algorithm(s) appropriately. Another way to avoid the need for a post-processing self-calibration is to include a set of appropriate
terms in the PAD08 modelling scheme.

We hypothesise that the sky over-subtraction problem discussed by AIH11 is the main cause of the detected systematic trend
in the aperture magnitudes as a function of PSF FWHM and object brightness, although there may also be some contribution to the
trend from aperture flux losses. With regards to the complicated systematic
trends that we have found in the PSF magnitudes as a function of PSF FWHM and object brightness, we do not speculate
on exactly what is causing them because we do not have access to sufficient information on the details of the PSF photometry routines in the
SDSS pipeline. Having studied the basic descriptions of these routines in \citet{lup2001} and \citet{sto2002}, we flag the following
procedures as potentially containing algorithms that introduce the detected systematic trends into the PSF magnitudes:
\begin{itemize}
\item{The PSF model adopted for each image including the model for the spatially variable terms.}
\item{The procedure for fitting the PSF model to each object in order to measure the PSF flux, which includes sinc-resampling of the image pixel data
      for the objects.}
\item{The procedure for determining the aperture correction that is applied to the PSF magnitude measurements.}
\end{itemize}

The systematic trends in the PSF magnitudes as a function of subpixel coordinates for the regime of under-sampled PSFs
are also most likely caused by some aspect(s) of the above procedures, and, specifically, we suspect that they are due to a slight mismatch between the
PSF model and the actual shape of the image PSF. We are unable to offer an explanation for the unexpected systematic trend as a function of subpixel coordinates for
normally-sampled and well-sampled PSFs that is present at the same amplitude in all wavebands for the PSF magnitudes, and which is detected
in the $r$ waveband for the aperture magnitudes.

We also found that the SDSS aperture and PSF magnitude scales are related by a non-linear transformation
that departs from linearity by $\sim$1-4\%, implying that at least one of the scales 
departs from a pure asinh-magnitude scale. Furthermore, this discovery invalidates to some extent the application of the PAD08 calibrations to the PSF
magnitudes, since the calibration model parameters are derived from fitting the aperture magnitudes. To avoid the problems introduced by the non-linear relation
between the magnitude scales, it would be necessary to either transform the aperture magnitudes to the PSF magnitude scale before performing the 
calibration model fit, or to transform the PSF magnitudes to the aperture magnitude scale before applying the photometric calibrations derived from the
aperture magnitudes, or, better still, to simply identify and remedy the cause of the non-linear relation in the first place. We have argued that the main cause
of the non-linear relation between the magnitude scales is most likely yet another consequence of sky over-subtraction in the SDSS images.
Finally, we report the detection of a $\sim$0.5-1\% zero-point difference between the aperture and PSF magnitude scales for the $u$ waveband.

Our results indicate that there is still room for improvement in the relative photometric calibration of the SDSS photometric data.
PAD08 report that they detect systematic trends in the residuals of the SDSS aperture magnitudes for the PAD08 model ``at the $\sim$0.5\%
level''. They also perform simulations of the data that include random walks in time of the atmospheric extinction coefficients to simulate temporal
atmospheric variations, and they fit the simulated data with the photometric calibration model to see if they can reproduce the systematic errors
at the $\sim$0.5\% level that they have detected in the real data. PAD08 find that their simulations produce systematic trends at the level
of $\sim$13, 8, 8, 7, and 8~mmag in the $ugriz$ wavebands, respectively, and therefore, they hypothesise that it is the unmodelled temporal atmospheric
variations in the PAD08 modelling scheme that are the cause of the remaining systematic trends in the SDSS photometry. However,
the results from their simulations depend heavily on the assumptions about the temporal atmospheric variations.
We have shown, in contrast, that there are still systematic trends not related to temporal atmospheric variations in the SDSS photometry at the
0.5\% level and above for both the aperture and PSF magnitudes. Hence, it is possible to further improve
the precision of the relative calibration of the SDSS photometric data, which, if implemented, will be of great benefit to the astronomical community.

The SDSS photometry is now routinely used to calibrate other photometric data, including other large scale surveys (e.g. \citealt{bro2011}, \citealt{ses2011}, etc.).
Specifically for this purpose, \citet{ive2007} have produced a standard star catalogue for Stripe 82 based on the SDSS repeat observations. Therefore, the
correction of the systematic trends and magnitude scale non-linearities that we have found in order to produce the most precise SDSS photometric data is crucial.

Finally, all of the issues that we have highlighted with the SDSS photometry are likely be relevant to other upcoming surveys (e.g.
PanSTARRS - \citealt{kai2002}, DES - \citealt{fla2005}, LSST - \citealt{ive2008}, etc.) in their analysis
of the photometric precision that they achieve. Our paper emphasises the fact that the relative calibration of the well-established SDSS photometry
could still be improved further for little effort, and that a thorough investigation and understanding of the systematic trends that are
present in the photometry is very important when constructing a photometric calibration model so as to include all of the relevant terms.

\section*{Acknowledgements}

DMB dedicates this work to those close friends who sadly passed away before their time: Nick Holliday 
- school form teacher and inspiration, and Gin\'es Ram{\'{i}}rez Alem\'an - full of fun and life.
DMB and WF also dedicate this work to a close collaborator and dear friend who also passed away recently, Carlo Izzo.

This research has made extensive use of the DanIDL library of routines (http://www.danidl.co.uk).
We would like to thank the SDSS helpdesk, especially Ani Thakar, for supplying sufficient space for the large CasJobs
queries that we submitted. Lodovico Coccato kindly tested in depth the program {\tt fit\_photometric\_calibration.pro} and its sub-modules.

Funding for SDSS-III has been provided by the Alfred P. Sloan Foundation, the Participating Institutions, the National Science Foundation,
and the U.S. Department of Energy Office of Science. The SDSS-III web site is http://www.sdss3.org/ where information on the 
Participating Institutions may be found.

\label{lastpage}

\end{document}